\pdfoutput=1
\documentclass[11pt,twoside,a4paper,cmspaper,final,collab]{cms-tdr}
\def\svnVersion{79ef0e2}\def\svnDate{2022/01/21}\def\cmsCernNoTag{CERN-EP-2021-102}\def\cmsCernDate{\today}\def\cmsMessage{Published in Physics Letters B as \href{http://dx.doi.org/10.1016/j.physletb.2022.136888}{\doi{10.1016/j.physletb.2022.136888}.}}
\begin{document}\cmsNoteHeader{EXO-20-001}

\newcommand{\Wg}{\ensuremath{\PW\PGg}\xspace}
\newcommand{\MJ}{\ensuremath{m_\mathrm{J}^\text{SD}}\xspace}
\newcommand{\etaJ}{\ensuremath{\eta_\mathrm{J}}\xspace}
\newcommand{\etag}{\ensuremath{\eta_\PGg}\xspace}
\newcommand{\MWg}{\ensuremath{m_{\Jg}}\xspace}
\newcommand{\Jg}{\ensuremath{\mathrm{J}\PGg}\xspace}
\newcommand{\tauto}{\ensuremath{\tau_{21}}\xspace}
\newcommand{\ptg}{\ensuremath{\pt^{\PGg}}\xspace}
\newcommand{\costh}{\ensuremath{\cos\theta^*_\PGg}\xspace}
\newlength\cmsTabSkip\setlength{\cmsTabSkip}{1ex}
\newcommand{\GX}{\ensuremath{\Gamma_\PX}\xspace}
\newcommand{\mX}{\ensuremath{m_\PX}\xspace}

\cmsNoteHeader{EXO-20-001} 
\title{Search for \texorpdfstring{$\Wg$}{Wgamma} resonances in proton-proton collisions at \texorpdfstring{$\sqrt{s} = 13\TeV$}{sqrt(s) = 13 TeV} using hadronic decays of Lorentz-boosted \texorpdfstring{$\PW$}{W} bosons}

\author*[inst1]{The CMS Collaboration}
\date{\today}

\abstract{
A search for \Wg resonances in the mass range between 0.7 and 6.0\TeV is presented. The \PW boson is reconstructed via its hadronic decays, with the final-state products forming a single large-radius jet, owing to a high Lorentz boost of the \PW boson. The search is based on proton-proton collision data at $\sqrt{s} = 13\TeV$, corresponding to an integrated luminosity of 137\fbinv, collected with the CMS detector at the  LHC in 2016--2018. The \Wg mass spectrum is parameterized with a smoothly falling background function and examined for the presence of resonance-like signals. No significant excess above the predicted background is observed. Model-specific upper limits at 95\% confidence level on the product of the cross section and branching fraction to the \Wg channel are set. Limits for narrow resonances and for resonances with an intrinsic width equal to 5\% of their mass, for spin-0 and spin-1 hypotheses, range between 0.17\unit{fb} at 6.0\TeV and 55\unit{fb} at 0.7\TeV. These are the most restrictive limits to date on the existence of such resonances over a large range of probed masses. In specific heavy scalar (vector) triplet benchmark models, narrow resonances with masses between 0.75 (1.15) and 1.40 (1.36) TeV are excluded for a range of model parameters. Model-independent limits on the product of the cross section, signal acceptance, and branching fraction to the \Wg channel are set for minimum \Wg mass thresholds between 1.5 and 8.0\TeV.}

\hypersetup{%
pdfauthor={CMS Collaboration},%
pdftitle={Search for Wgamma resonances in proton-proton collisions at sqrt(s)= 13 TeV using hadronic decays of Lorentz-boosted W bosons},%
pdfsubject={CMS},%
pdfkeywords={CMS, Wgamma resonances, BSM particles}}

\maketitle 

\section{Introduction}

Searches for new resonances predicted in theories beyond the standard model (SM) are among the key components of the physics program at the CERN LHC. Many of these searches have been carried out in the past decade, since the start of the LHC operation, and have helped to reshape the landscape of allowed beyond-the-SM physics. Experimental searches for the production of new resonances decaying into a pair of SM particles are a notable aspect of this program. In the case of a new resonance with significant couplings to quarks and gluons, it would be produced in quark-antiquark, quark-gluon, or gluon-gluon interactions in proton-proton ($\Pp\Pp$) collisions at the LHC and subsequently decay into a pair of jets. On the other hand, if the couplings to quarks and gluons were suppressed, other decay channels, including decays into pairs of vector bosons (diboson decays), would become dominant.

Diboson decays involving photons, \ie, $\Wg$, $\PZ\Pgg$, $\PH\PGg$, and $\PGg\PGg$ channels, are important parts of this search program, owing to the excellent detection efficiency and photon energy resolution of the ATLAS and CMS detectors. For example, searches in the $\PGg\PGg$ channel contributed significantly to the discovery of the Higgs boson by the ATLAS and CMS Collaborations in 2012~\cite{Higgs1,Higgs2,Higgs3}. Nevertheless, the above-mentioned decay modes involving photons, in particular the \Wg decay mode, are generally less studied over a large range of masses than other diboson signatures. There are multiple beyond-the-SM theories that predict \Wg resonances, including new particles in models with an extended Higgs sector~\cite{Gunion:1987ke}, such as charged Higgs bosons in generic two Higgs doublet models~\cite{2HDM,CapdequiPeyranere:1990qk}, as well as particles predicted in technicolor~\cite{TC1,TC2,TC3,TC4}, heavy vector triplet~\cite{HVT}, and electroweak singlet~\cite{Howe:2016mfq} models, or scalar ``quirks" in folded supersymmetry~\cite{Burdman:2008ek}. A number of nonresonant \Wg analyses that mainly probe anomalous $\PW\PW\PGg$ couplings have been conducted at the CERN S$\Pp\PAp$S, LEP, and LHC, as well as at the Fermilab Tevatron. However, searches for \Wg resonances at high mass have been conducted only by the ATLAS experiment, in the leptonic decay channel of the \PW\ boson at the center-of-mass energies of 7\TeV~\cite{ATLAS-Wg7} and 8\TeV~\cite{ATLAS-Wg8}, and in the hadronic decay channel at 13\TeV~\cite{ATLAS-Wg13}. The best 95\% confidence level (\CL) upper limits on the product of the cross section and branching fraction to \Wg for narrow spin-1 resonances in the leptonic channel are 6.0--0.5\unit{fb} in the 0.2--1.6\TeV mass range~\cite{ATLAS-Wg8}, while the analogous limits in the hadronic channel~\cite{ATLAS-Wg13} vary between 12 and 0.14\unit{fb} within the 1.0--6.8\TeV mass range probed.

In this Letter, we describe a search for \Wg resonances in the hadronic decay channel of the \PW\ boson using $\Pp\Pp$ collision data at $\sqrt{s} = 13\TeV$ delivered by the LHC in 2016--2018. The results of the search are interpreted in terms of limits on narrow and broad, spin-0 and spin-1 resonances in a mass range between 0.7 and 6.0\TeV. Narrow resonances are taken to be those with widths $\GX$  that are negligible compared to the experimental resolution, while for broad resonances we consider the representative case for which $\GX/\mX = 5\%$, where $\mX$ is the resonance mass. Given the large mass of the resonances probed in this analysis, the \PW\ boson is produced with a high Lorentz boost and is reconstructed as a single large-radius jet. The two-prong structure and mass of this jet are established using jet substructure techniques, allowing for the reduction of the dominant background from direct photon production, where a jet recoiling against a photon originates from quantum chromodynamics (QCD) radiation.

Digitized versions of tables and plots from this paper can be found in the HEPData database~\cite{hepdata}.

\section{The CMS detector}

The central feature of the CMS apparatus is a superconducting solenoid of 6\unit{m} internal diameter, providing a magnetic field of 3.8\unit{T}. Within the solenoid volume are a silicon pixel and strip tracker, a lead tungstate crystal electromagnetic calorimeter (ECAL), and a brass and scintillator hadron calorimeter (HCAL), each composed of a barrel and two endcap sections. Forward calorimeters extend the pseudorapidity ($\eta$) coverage provided by the barrel and endcap detectors. Muons are detected in gas-ionization chambers embedded in the steel flux-return yoke outside the solenoid.

In the region $\abs{\eta} < 1.74$, the HCAL cells have widths of 0.087 in $\eta$ and 0.087 in azimuth ($\phi$). In the $\eta$-$\phi$ plane, and for $\abs{\eta} < 1.48$, the HCAL cells map on to $5{\times}5$ arrays of ECAL crystals to form calorimeter towers projecting radially outwards from close to the nominal interaction point. For $\abs{\eta } > 1.74$, the coverage of the towers increases progressively to a maximum of 0.174 in $\Delta \eta$ and $\Delta \phi$. Within each tower, the energy deposits in ECAL and HCAL cells are summed to define the calorimeter tower energies, and subsequently used to provide the energies and directions of hadronic jets.

Events of interest are selected using a two-tiered trigger system~\cite{Khachatryan:2016bia}. The first level (L1), composed of custom hardware processors, uses information from the calorimeters and muon detectors to select events at a rate of around 100\unit{kHz} within a fixed latency of about 4\mus~\cite{Sirunyan:2020zal}. The second level, known as the high-level trigger, consists of a farm of processors running a version of the full event reconstruction software optimized for fast processing, and reduces the event rate to around 1\unit{kHz} before data storage~\cite{Khachatryan:2016bia}.

A more detailed description of the CMS detector, together with a definition of the coordinate system used and the relevant kinematic variables, can be found in Ref.~\cite{Chatrchyan:2008zzk}.

\section{Data sets and event selection}
\label{sec:selection}

The data used in this search correspond to an total integrated luminosity of 137\fbinv and were recorded by the CMS experiment at $\sqrt{s} = 13$\TeV in 2016--2018 (36, 41, and 60\fbinv in 2016, 2017, and 2018, respectively)~\cite{LUM-17-001, LUM-17-004, LUM-18-001}. The high instantaneous luminosity delivered by the LHC results in additional interactions in the same or neighboring bunch crossings as the hard scattering interactions (pileup). The average number of pileup interactions in the 2016 (2017 and 2018) data set is around 23 (32).

The data are selected via single-photon triggers that require the photon candidate to be within $\abs{\eta} < 2.5$, and to have transverse momentum $\pt > 165$ or 175\GeV in 2016 and $\pt > 200$\GeV in 2017--2018. We determine the selection efficiencies for these triggers
using unbiased data samples collected with single-muon triggers. The single-photon triggers are found to be 98--100\% efficient with respect to the offline selection described below, for the entire mass range used in the analysis. The small residual inefficiency is taken into account when calculating the signal acceptance.

Simulated Monte Carlo (MC) signal samples are produced at leading order (LO) in perturbative QCD. They are used to optimize the analysis selection and to calculate the signal efficiency. Simulated signal events of spin-0 resonances decaying to \Wg are generated using electroweak triplet pseudo-Goldstone bosons $\PGp_3$~\cite{Arkani-Hamed:2016kpz} and a heavy (pseudo-)scalar $SU(2)_\mathrm{L}$ triplet~\cite{Capdevilla:2019zbx}, while for spin-1 resonances, a heavy vector $SU(2)_\mathrm{L}$ triplet~\cite{Capdevilla:2019zbx} is used. Several signal samples are generated with masses ranging from 0.7 to 6.0\TeV. Two resonance width assumptions are used in the simulation: one, termed ``narrow", has a width which is significantly smaller than the detector resolution that ranges between 3.75 and 5\%, and the second, referred to as ``broad", has $\GX/\mX = 5\%$. The latter choice is representative of broad resonances, for which the impact of the off-shell production on the signal efficiency becomes sizable.

Simulated background events do not enter the analyses directly, as the background is obtained from a fit to
data. They are only used to assess the accuracy of the background model and to optimize the event selection. The dominant background from $\Pgg$+jet production as well as the QCD multijet background from SM events composed uniquely of jets produced through the strong interaction which have a jet misidentified as a photon are generated at LO using \MGvATNLO. Smaller backgrounds from $\PW$+jets and $\PW$+\PGg production, as well as top quark backgrounds, are not simulated, as their contribution is far less than that of the dominant backgrounds and does not affect the search optimization procedure.

Both the signal and background samples are generated using \MGvATNLO 2.2.2 (2.4.2)~\cite{aMCNLO} with NNPDF3.0 NLO~\cite{Ball:2014uwa} (NNPDF3.1 NNLO~\cite{Ball:2017nwa}) parton distribution functions (PDFs) for 2016 (2017 and 2018) conditions. Fragmentation and hadronization are simulated with \PYTHIA 8.205 (8.230)~\cite{Sjostrand:2014zea} with the CUETP8M1~\cite{Monash,CUETP8M1} (CP5~\cite{Sirunyan:2019dfx}) underlying event tune for 2016 (2017 and 2018) samples. All simulated samples are processed with the full CMS detector model based on \GEANTfour~\cite{Geant} and reconstructed with the same suite of programs as used for collision data. Pileup effects are taken into account by superimposing simulated minimum bias events on the hard scattering interaction, with the multiplicity distribution matching that observed in data.

A particle-flow (PF) event algorithm~\cite{PFlow} is used, which aims to reconstruct and identify each individual particle in an event, with an optimized combination of information from the various elements of the CMS detector. The energy of photons is obtained from the ECAL measurement. The energy of electrons is determined from a combination of the electron momentum at the primary interaction vertex as determined by the tracker, the energy of the corresponding ECAL cluster, and the energy sum of all bremsstrahlung photons spatially compatible with originating from the electron track. The energy of muons is obtained from the curvature of the corresponding track. The energy of charged hadrons is determined from a combination of their momentum measured in the tracker and the matching ECAL and HCAL energy deposits, corrected for the response function of the calorimeters to hadronic showers. Finally, the energy of neutral hadrons is obtained from the corresponding corrected ECAL and HCAL energies.

The events must contain at least one reconstructed primary vertex with at least four associated tracks, with transverse (longitudinal) coordinates required to be within 2\,(24)\unit{cm} of the nominal collision point. The candidate vertex with the largest value of summed physics-object $\pt^2$ is taken to be the primary $\Pp\Pp$ interaction vertex. The physics objects are the jets, clustered using the jet finding algorithm~\cite{Cacciari:2008gp,Cacciari:2011ma} with the tracks assigned to candidate vertices as inputs, and the associated missing transverse momentum, taken as the negative vector sum of the \pt of those jets.

Since the dominant background in the analysis is from direct photon production ($\Pgg$+jets), rather than from sources with a misidentified photon, we chose a ``loose" photon identification working point of a standard CMS sequential-selection algorithm, which maximizes the photon efficiency at the cost of a slightly higher misidentification rate compared to other available working points~\cite{CMS-EGM-14-001}. The identification is based on photon shower shape and isolation variables. The latter are computed from various types of PF candidates in a cone of radius $\Delta R = \sqrt{\smash[b]{(\Delta\eta)^2+(\Delta\phi)^2}} = 0.3$ around the photon candidate, corrected for the pileup effects. In addition, a conversion-safe electron veto~\cite{CMS-EGM-14-001} is applied. The loose working point gives an efficiency of approximately 90\% that does not depend on the photon \pt up to the highest values explored in the analysis, while reducing the background from misidentified photons by approximately a factor of 7. The photon candidates are required to have $\pt > 225$\GeV and to be within the barrel fiducial region of the ECAL ($\abs{\eta} < 1.44$). Since events with a photon reconstructed in the endcap region suffer from high background from $\Pgg$+jets, which peaks in the forward direction, they do not add to the sensitivity of the analysis and therefore are not included.

Large-radius jets ($\mathrm{J}$) are used to reconstruct hadronically decaying, highly Lorentz-boosted $\PW$ boson candidates. These jets are reconstructed from PF candidates clustered using the anti-\kt
algorithm~\cite{Cacciari:2008gp,Cacciari:2011ma} with a distance parameter of 0.8. Charged hadrons not originating from the primary vertex are not considered in the jet clustering. The pileup per particle identification algorithm (PUPPI)~\cite{Sirunyan:2020foa,Bertolini:2014bba} is used to mitigate the effect of pileup at the reconstructed particle level, making use of local energy distribution, event pileup properties, and tracking information. Charged particles identified to be originating from pileup vertices are discarded. The momenta of the neutral particles are rescaled according to their probability to originate from the primary interaction vertex deduced from the local shape variable, superseding the need for jet-based pileup corrections~\cite{Sirunyan:2020foa}. Jet energy corrections are derived from simulation studies so that the average measured energy of jets becomes identical to that of particle-level jets. In situ measurements of the momentum balance in dijet, $\Pgg$+jet, $\PZ$+jet, and multijet events are used to determine any residual differences between the jet energy scale in data and in simulation, and appropriate corrections are applied~\cite{Khachatryan:2016kdb}. Additional quality criteria~\cite{CMS-PAS-JME-16-003} are used to remove jets due to rare spurious noise patterns in the calorimeters, and also to suppress leptons misidentified as jets. The jet energy resolution typically amounts to 15\% at 10\GeV, 8\% at 100\GeV, and 4\% at 1\TeV. In each event, the jet selected to be the hadronic \PW candidate must have $\pt > 225$\GeV, to balance the $\pt$ of the selected photon, and is required to be separated from the photon by a distance of $\Delta R > 1.1$ to reduce the contamination of the photon isolation cone with the jet constituents.

Since the signal jets are merged products of the \PW\ boson decay, we require the jet mass to be within a certain range of the \PW\ boson mass to reduce the very large background from QCD jets, which have steeply falling jet mass distribution. To improve the signal and background separation, a jet grooming algorithm known as ``soft drop" (SD)~\cite{softdrop}, with parameters $\beta=0$ and $z_\text{cut}=0.1$, is applied to recursively remove soft, wide-angle radiation from anti-\kt jets. The groomed jet mass (\MJ) is then computed from the four-momentum sum of the remaining jet constituents, whose energies are corrected with the same factor that has already been used in the generic jet reconstruction described above. The typical mass resolution for a \PW\ boson jet is 10\%~\cite{Sirunyan:2017isc}. Finally, in order to avoid the region of rapidly varying efficiency near threshold, we require the invariant mass of the selected jet and photon (\MWg) to exceed 0.6\TeV.

\section{Analysis optimization}
\label{sec:optimization}

The analysis is optimized using a sequential selection on a number of kinematic variables. These variables fall into two classes: those related to the resonance decay kinematics and those related to the properties of a large-radius jet. The former are: pseudorapidities of the photon \etag and the jet \etaJ, the cosine of the polar decay angle in the center-of-mass frame of the \Jg system with respect to the beam axis \costh, and the ratio of the photon transverse momentum \ptg to \MWg. The last two variables mentioned are highly correlated and are both used to separate the mostly central $s$-channel signal events from the mostly forward $t$-channel direct photon background. For the purpose of tagging jets originating from the \PW boson decay (\PW jet tagging), selections are also applied on two variables related to the properties of a large-radius jet, which are the jet mass \MJ and the jet substructure variable \tauto. \MJ peaks at the \PW mass for the signal and falls rapidly for the QCD background. The variable $\tauto$ is defined as the ratio of $\tau_2$ to $\tau_1$, where $\tau_N$ is a set of $N$-subjettiness~\cite{Nsubjettiness} variables. Such variables are measures of how likely it is that a large-radius jet has a substructure of $N$ subjets. For a jet with exactly $N$ subjets, $\tau_N$ tends to small values, while $\tau_M$ values for $M \neq N$ are shifted to larger values. Thus, $\tauto$ is expected to be generally smaller for a signal, which contains a jet produced by an overlap of the quark jets from a two-prong \PW boson decay, than for the background, which mostly consists of structureless QCD jets.

The optimization aims at maximizing the expected signal significance for a large range of tested masses, where the background yield ($B$) is typically much larger than that of the signal we probe ($S$), and is large enough to use the Gaussian approximation, \ie, maximizing $S/\sqrt{B}$. This figure of merit (FOM) has the advantage that its maximum is independent of the signal cross section. The optimization is based on the events in the control region (CR) in data, defined as the lower sideband of the \PW boson jet mass $40 < \MJ < 65$ GeV, which has negligible signal contamination for the range of the signal cross sections probed in this analysis, as well as events in the signal region (SR) in signal simulation samples, which is defined as $68 < \MJ < 94\GeV$, as determined via the optimization using the FOM described above. The CR and SR are illustrated in Fig.~\ref{fig:regions}, which also shows the expected distributions in \MJ for the benchmark narrow spin-0 signals. It was demonstrated that the differences in this distribution between different signal spin and width hypotheses are negligible.

\begin{linenomath}
\begin{figure}[htb!]
\centering
\includegraphics[width=0.49\textwidth]{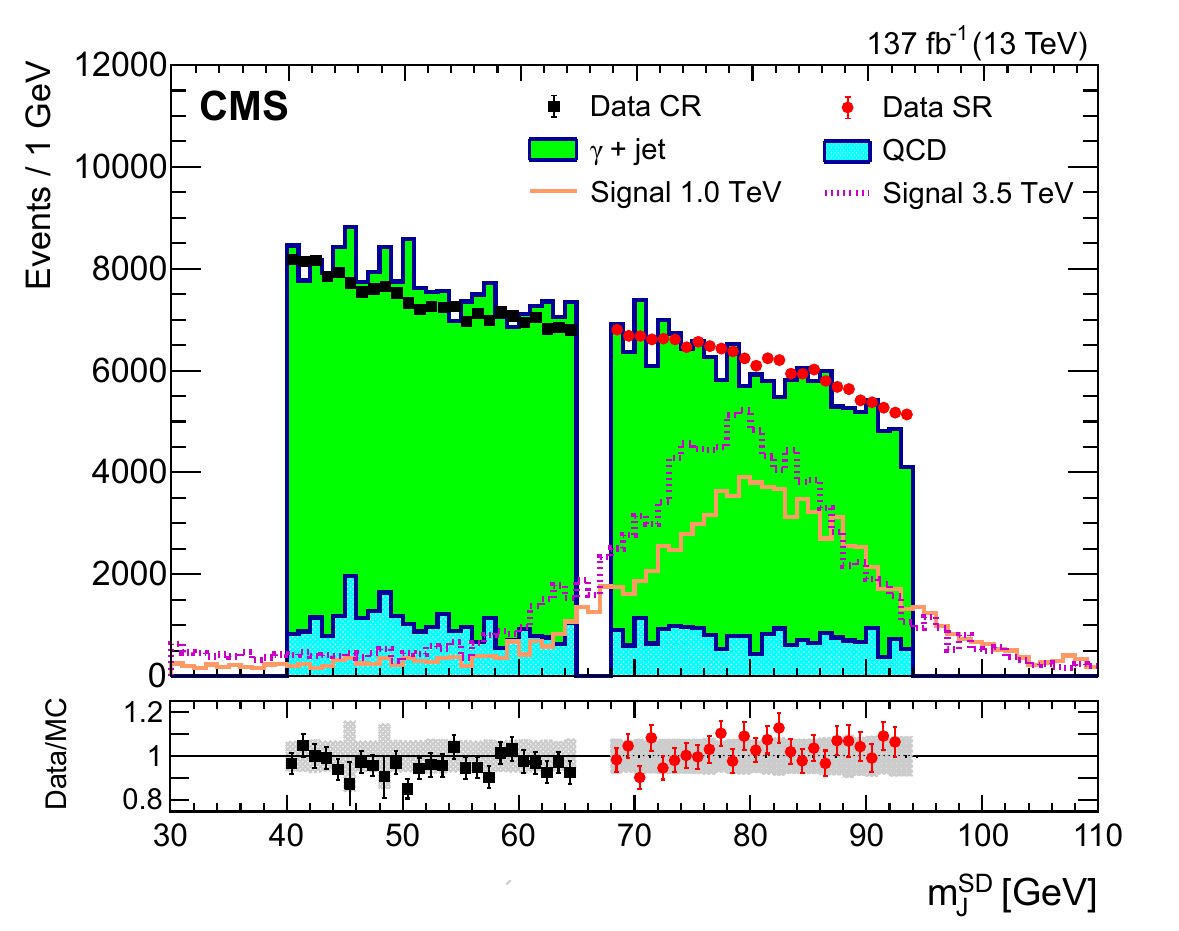}
\caption{Definitions of the signal and control regions in data, based on the jet mass \MJ. The stacked filled histograms represent dominant backgrounds from simulation, normalized to the $\ptg$ spectrum in the signal region. The red circles (black squares) correspond to data in the signal (control) region. Benchmark narrow spin-0 signal distributions, normalized to a cross section of 2\unit{pb} for two masses, 1.0 and 3.5 TeV, are shown by the solid orange and dashed magenta lines, respectively. The lower panel shows the data-to-simulation ratio in the control and signal regions. The gray hatched band shows the statistical uncertainty in the background estimation, stemming from the limited size of the simulated samples.}
\label{fig:regions}
\end{figure}
\end{linenomath}

The optimization of the FOM uses simulated signal events and two different estimates of the background in the signal region. The first background estimate uses data events in the CR, which are expected to have similar distributions in the main analysis variables to those in the SR. To ensure the same normalization, a scale factor of approximately 0.86 is applied to the CR event yield to match that in the SR. This procedure was justified by comparing the kinematic distributions of simulated background events in the SR and CR and finding that they are quite similar. The second background sample uses simulated background events that have been rescaled to account for the missing higher-order effects, as well as for an imperfect description of the misidentification of jets as photons in the QCD multijet simulation.  This is done by fitting the $\ptg$ spectrum in data with the sum of these two backgrounds, with the normalizations allowed to float in the fit. Figure~\ref{fig:data-MC} shows that before the optimization an adequate description of the SR data is achieved in the main kinematic variables used for the analysis optimization by using either MC simulation or CR data, with the exception of the $\tauto$ variable, which has a different shape in the data CR due to a strong correlation of the $\tauto$ variable with \MJ. Consequently, the optimization of the $\tauto$ selection is performed only using simulated backgrounds. The residual discrepancies seen in other distributions are due to missing minor backgrounds and the limitations of the MC simulation modeling of the data. They are typically present near the kinematic limits of the corresponding variables, far from the region of optimal selections, and consequently do not bias the optimization procedure. Since neither the simulated backgrounds nor data CR events are used in the final analysis, beyond the optimization step, we find that the level of agreement between the data and simulated backgrounds is adequate. Figure~\ref{fig:data-MC} also shows several benchmark signal points and, given the similarity of signal shapes for various spin and width combinations, the spin-0, narrow-width hypothesis is used, unless indicated otherwise in the legend.

\begin{linenomath}
\begin{figure*}[htbp!]
\centering
\includegraphics[width=0.42\textwidth]{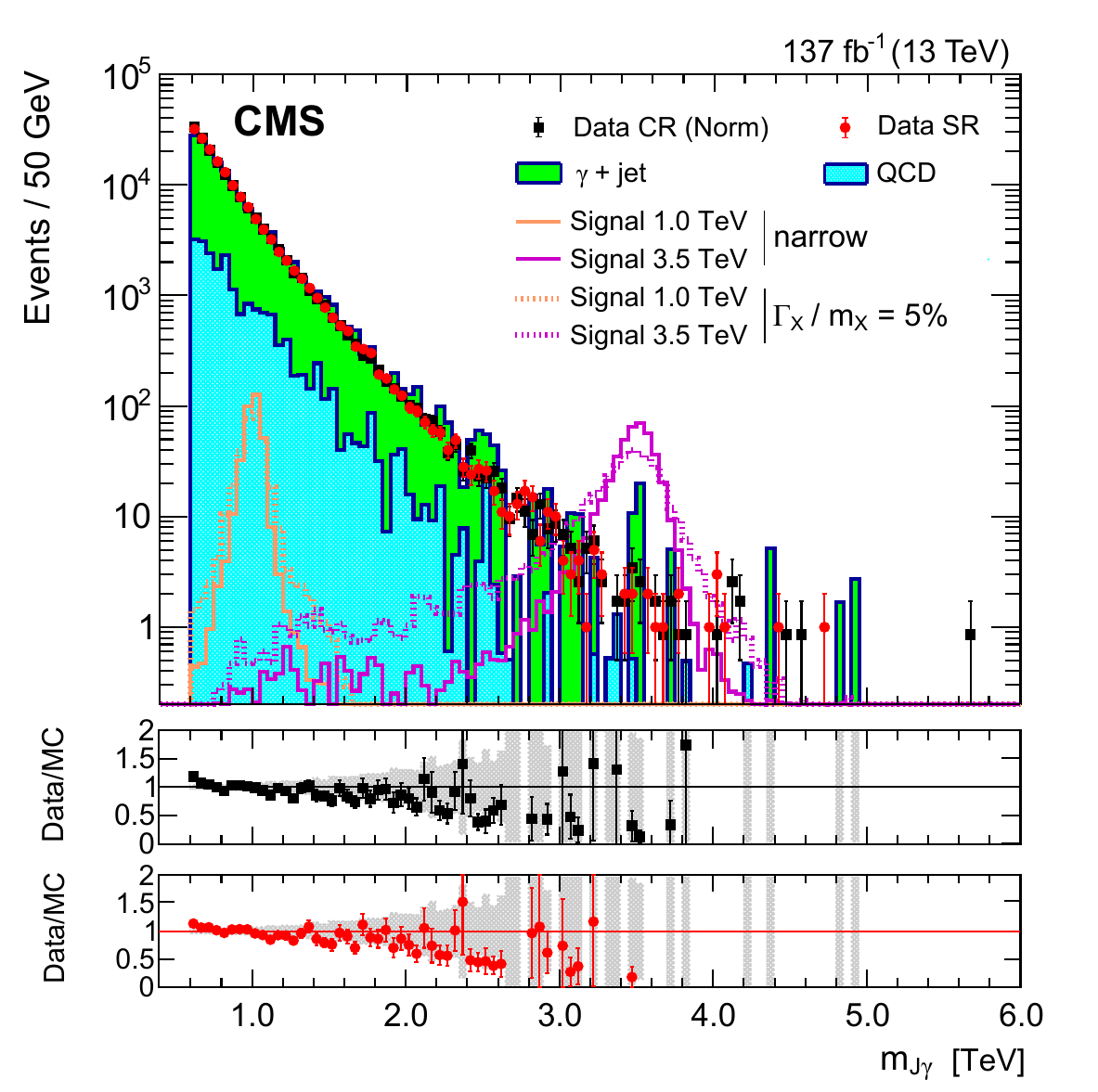}
\includegraphics[width=0.42\textwidth]{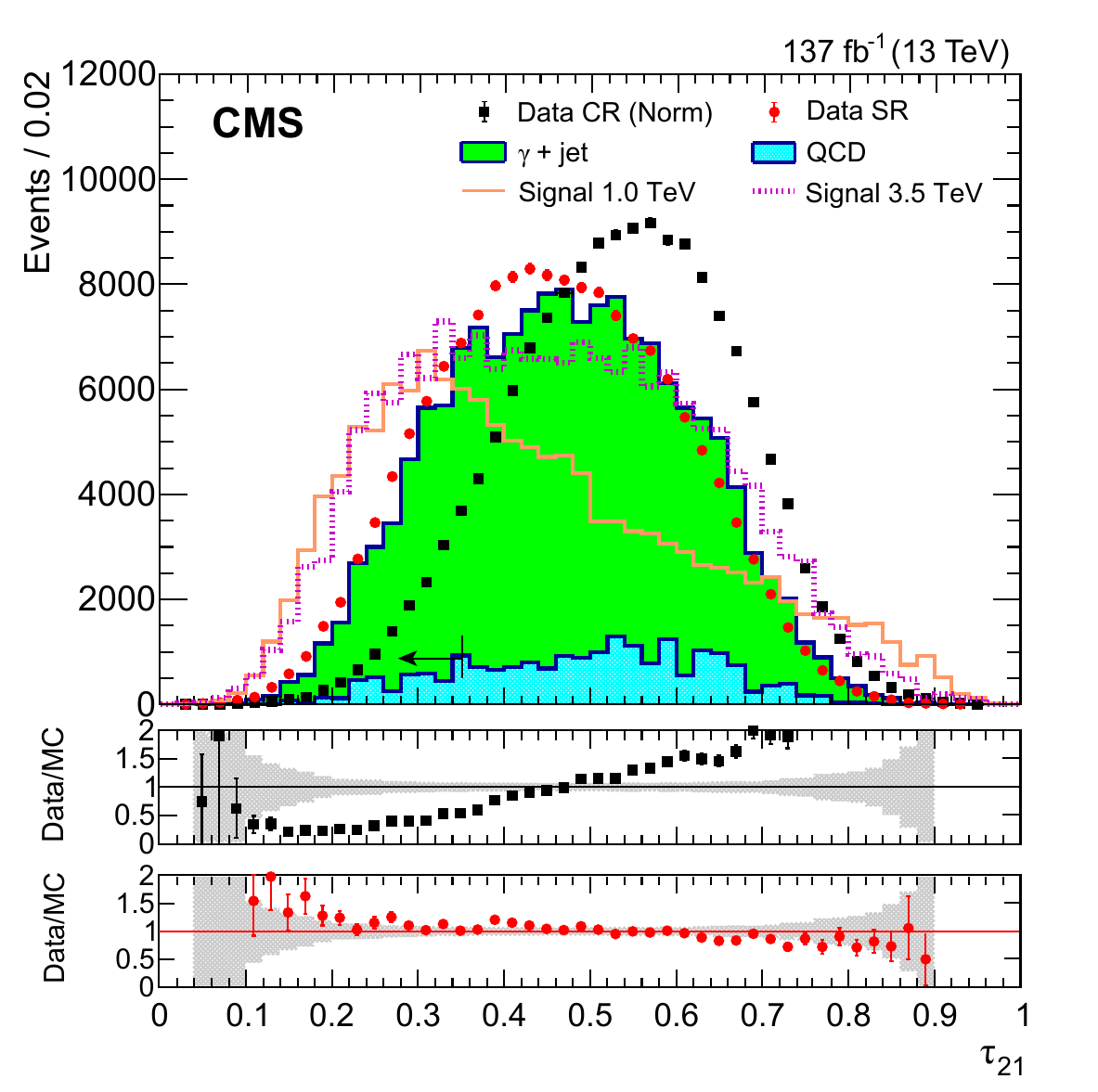}
\includegraphics[width=0.42\textwidth]{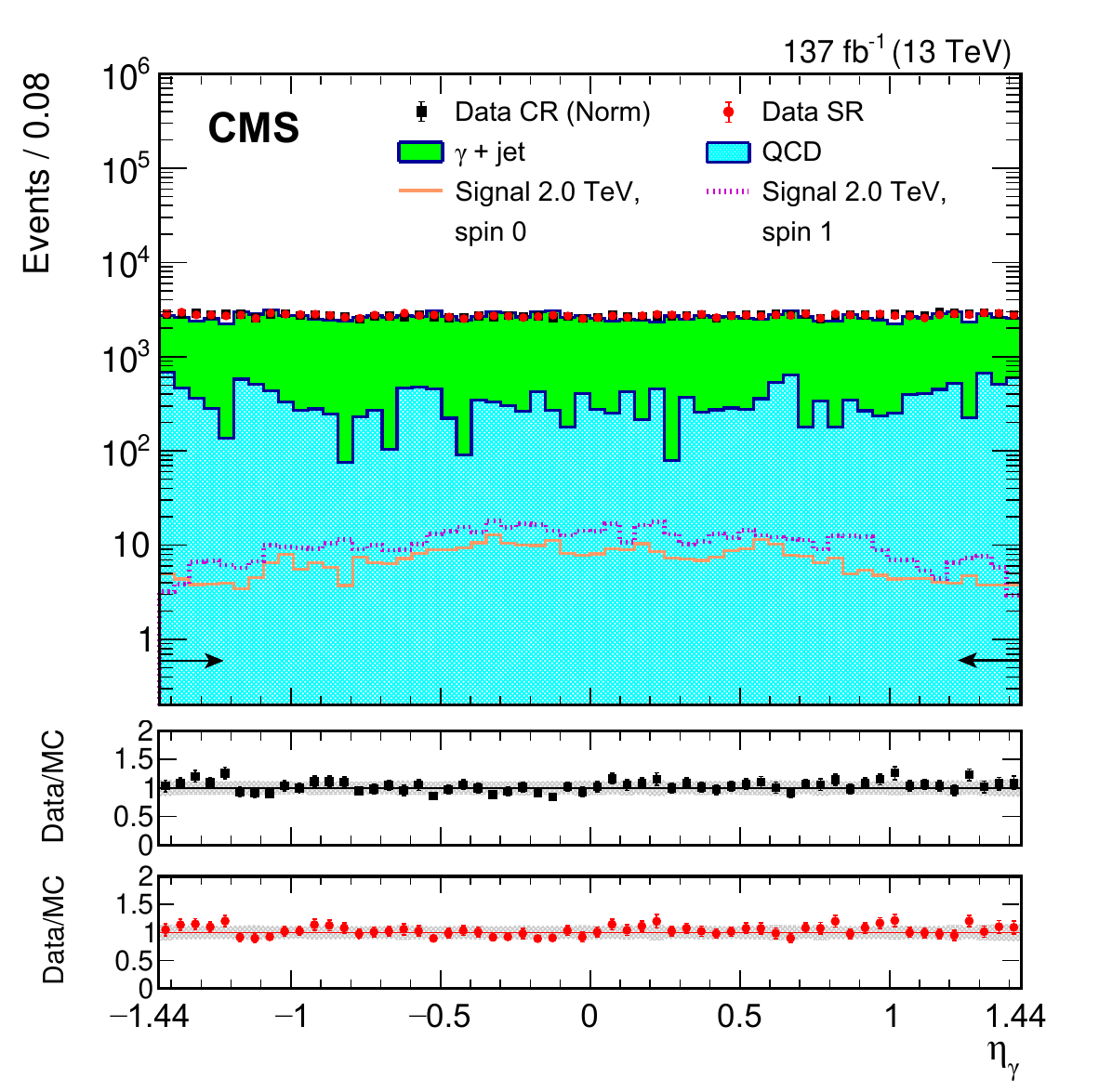}
\includegraphics[width=0.42\textwidth]{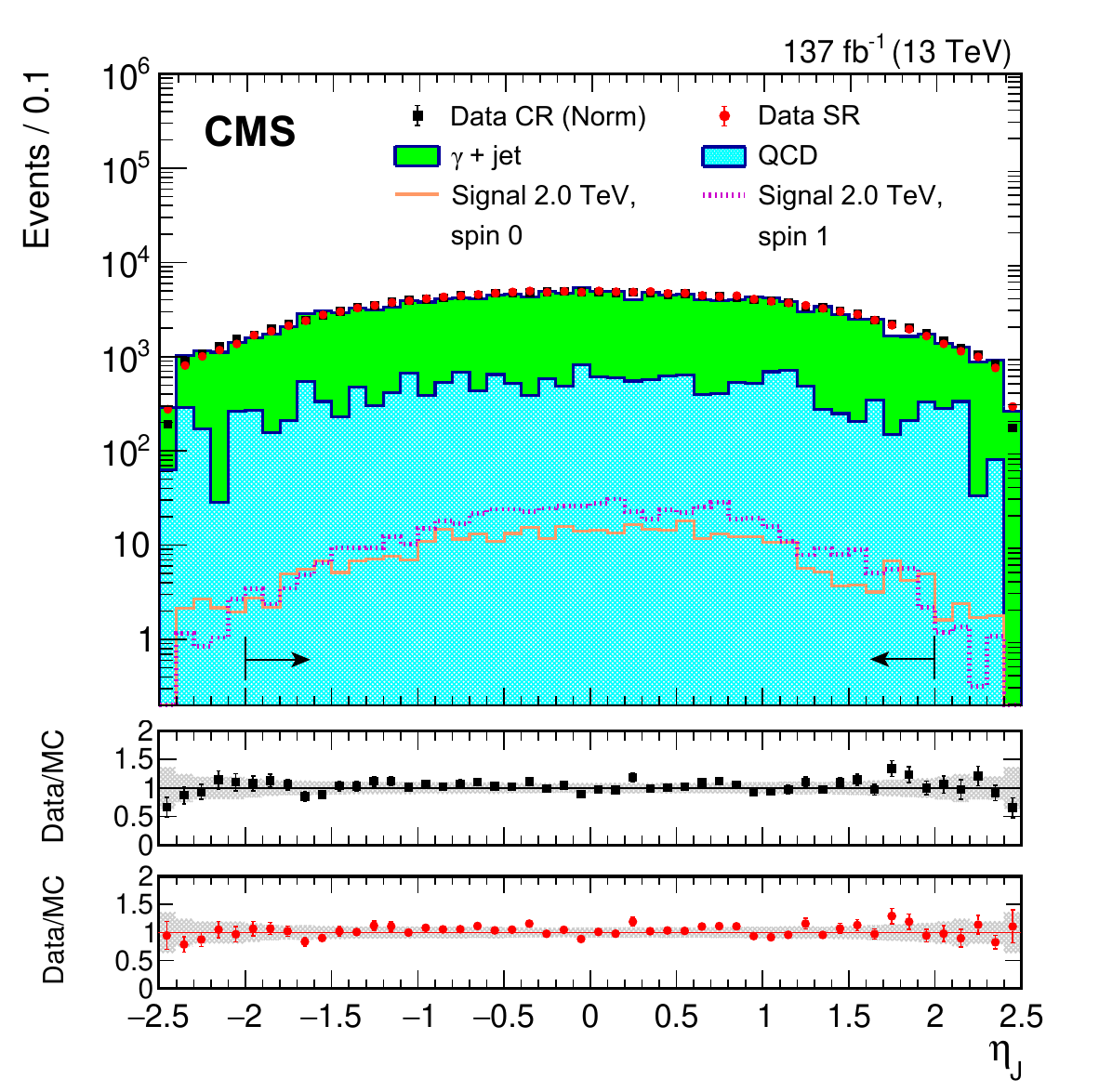}
\includegraphics[width=0.42\textwidth]{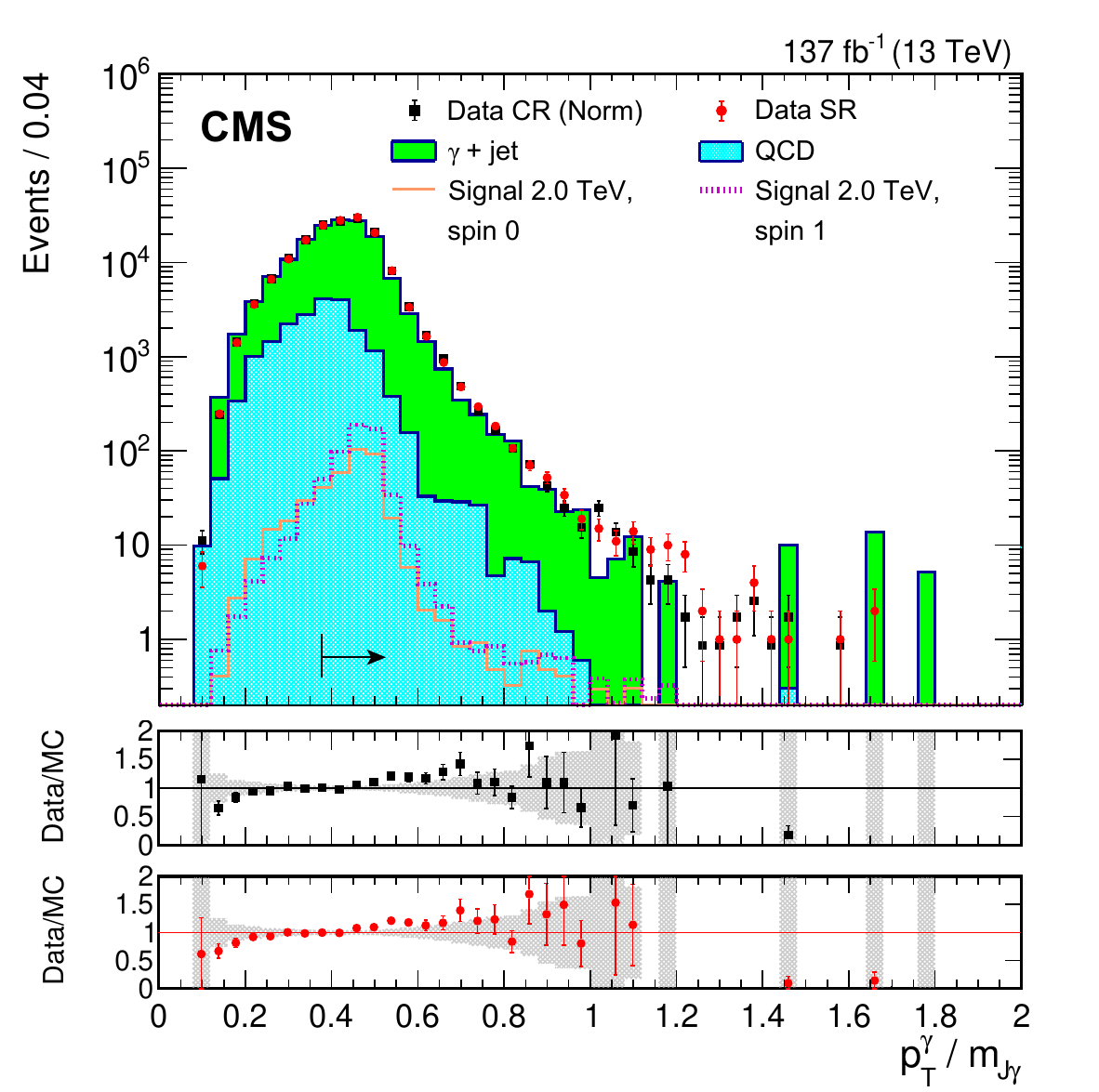}
\includegraphics[width=0.42\textwidth]{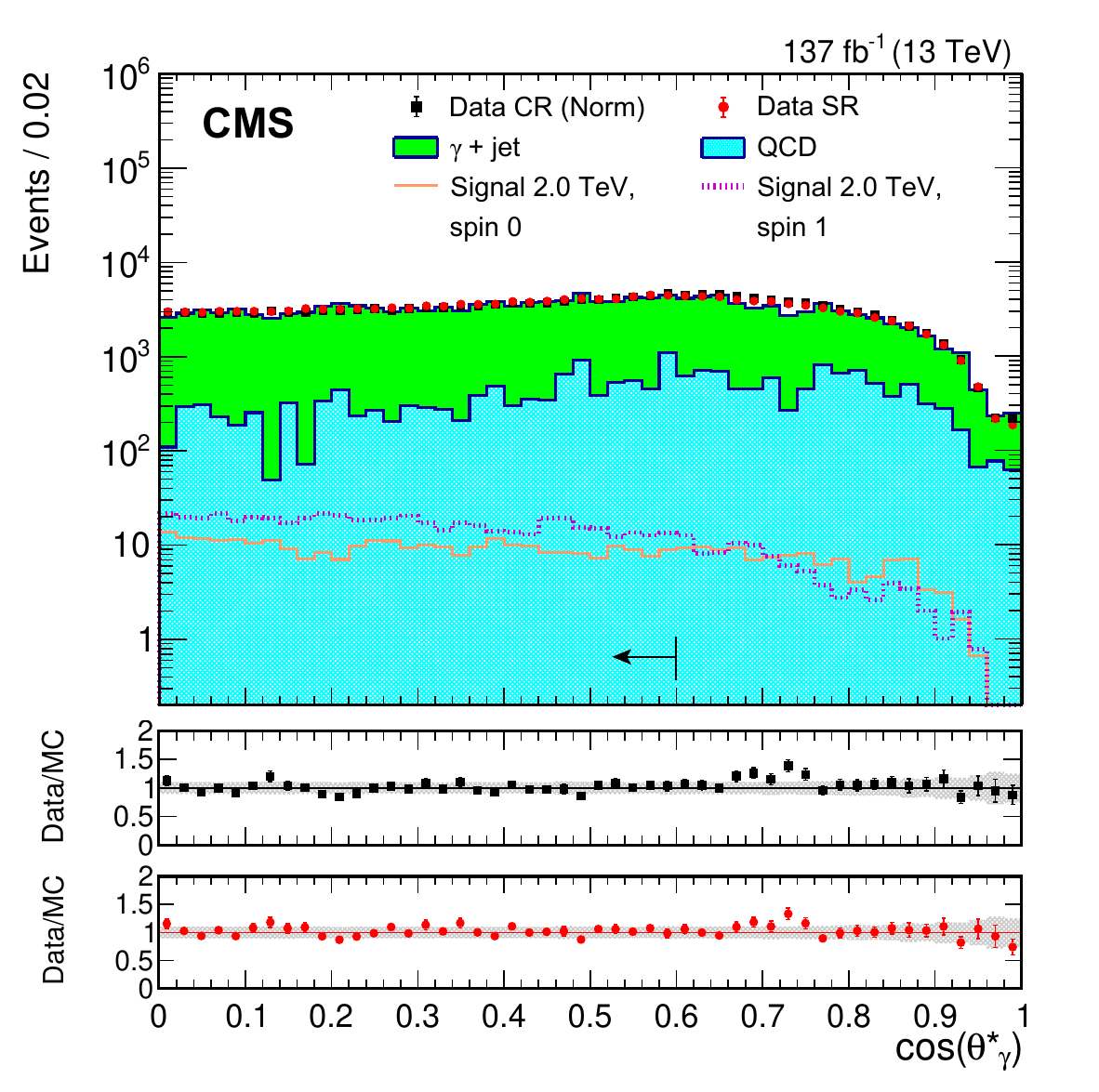}
\caption{Distributions of some of the kinematic variables used in the analysis. Upper row: \MWg (left), \tauto (right); middle row: $\etag$ (left), $\etaJ$ (right); lower row: $\ptg/\MWg$ (left), \costh (right), except that the yield in the control region is normalized to that in the signal region. Several benchmark signals are also shown, as indicated by the legend. By default, the spin-0, narrow width hypothesis is used unless indicated otherwise. Signals are normalized to a cross section of
5\unit{fb}, except for the \tauto distribution, for which the normalization is 2\unit{pb}. Optimized selections are indicated with the black arrows. The two lower panels show the data-to-simulation ratio in the control and signal regions, respectively. The gray hatched band shows the statistical uncertainty in the background estimation, stemming from the limited size of the simulated samples.}
\label{fig:data-MC}
\end{figure*}
\end{linenomath}

For most of the variables, and the signal masses, widths, and spins probed, the maxima of the FOM distributions are fairly broad, which justifies using a single set of selections for all signal masses and spin/width hypotheses, without compromising the search performance. We have tested whether combining the input variables into a multivariate discriminant using a boosted decision tree with adaptive boosting, instead of selecting on individual variables, results in a gain in performance. However, the best gain we were able to achieve was only a 5\% increase in the FOM value, which required separate discriminants constructed for each signal mass point. Consequently, a single set of selections on the individual variables was chosen, which simplifies the analysis without introducing a noticeable performance loss. The best selections chosen as a result of the above procedure are: $\abs{\etag} < 1.44$ (\ie, the photon in the ECAL barrel), $\abs{\etaJ} < 2.0$, $\tauto < 0.35$, $68 < \MJ < 94\GeV$, $\ptg/\MWg > 0.37$, and $\costh < 0.6$. The use of these optimal selections combined improves the FOM by up to 90\% for signals with masses ranging from 0.7\TeV to 3.5\TeV. The requirement on the $\tauto$ variable alone improves the FOM by 40\%, making it the most discriminating variable.

\section{Signal and background modeling}
\label{sec:modeling}

We describe the shape of the \MWg distributions for the signals probed by fitting simulated signal samples with analytic functions. For narrow resonances, we use a sum of Crystal Ball (CB)~\cite{CB} and Gaussian functions with different means. For broad resonances, we use a sum of CB and two Gaussian functions, where the two Gaussian functions have a common mean that may be different from that of the CB function. In order to obtain the signal shapes for the mass points where no simulated samples were generated, we use linear morphing between the adjacent simulated signal points~\cite{morphing}. The simulated signal samples for each of the three years of data taking reflect changes in MC tunes, pileup and trigger conditions, selection criteria, and detector performance along with time. We verify that the signal shape in simulation is consistent for the three years and use samples produced with the 2017 conditions as the signal model for all three data-taking years. Consequently, we also combine the \MWg spectra for the three data-taking periods and search for the presence of signal in this combined data set.

The overall signal acceptance $\mathcal{A}$, and the product of the acceptance and signal efficiency $\mathcal{A} \varepsilon$ for the optimal selection for spin-0 and spin-1 resonances, are shown as functions of the resonance mass in Fig.~\ref{fig:eff} (upper row). The latter ranges between 6.1 (10.0)\% and 12.3 (16.4)\% for spin-0 (spin-1) signals over the mass range probed. It was verified that for both $\mathcal{A}$ and $\mathcal{A} \varepsilon$ the differences between three years of data taking are small; thus the values for the 2017 data taking are used as the nominal ones. The efficiencies for narrow resonances are generally 1--2\% higher than for the broad ones, mainly because of the long low-mass tail in the \MWg distribution for broad resonances, caused by the quickly falling PDFs convoluted with the Breit--Wigner resonant shape. In order to be less sensitive to the exact description of this tail, which depends on both the PDF choice and the parameterization of the signal resonance line-shape, we use a window of $\pm 25\%$ of the resonance mass, centered on the mass. The size of the window corresponds to roughly $\pm 5$ effective widths of a broad resonance. The window requirement is included in the definition of the signal acceptance. The \PW jet tagging efficiency ($\epsilon_{\PW\text{-tag}}$, which is a part of the overall efficiency) is shown in Fig.~\ref{fig:eff} (lower), which illustrate that a slight decrease in the overall product of the acceptance and efficiency at high masses is due to a less effective \PW tagging for very energetic jets.
\begin{linenomath}
\begin{figure*}[tb!]
\centering
\includegraphics[width=0.49\textwidth]{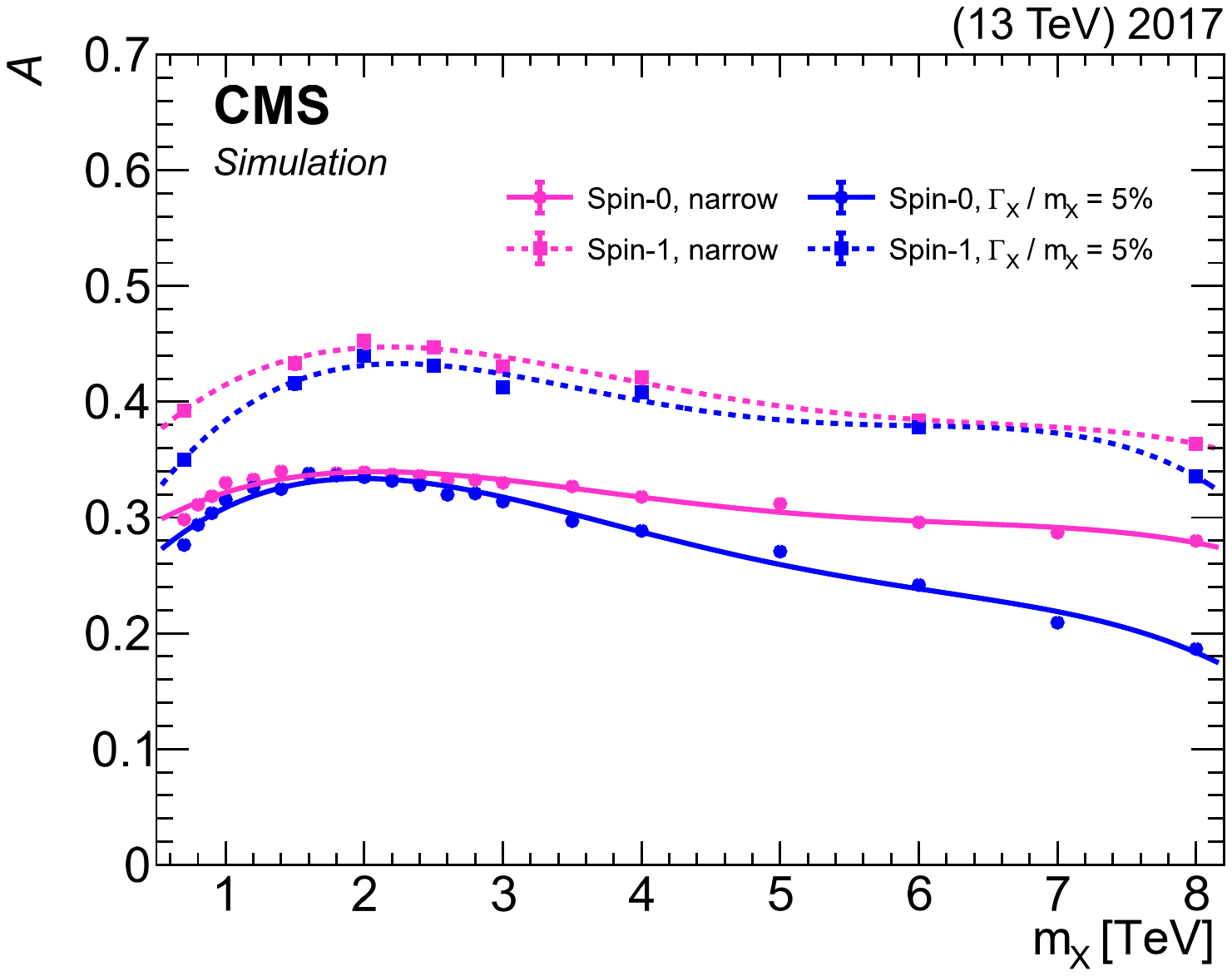}
\includegraphics[width=0.49\textwidth]{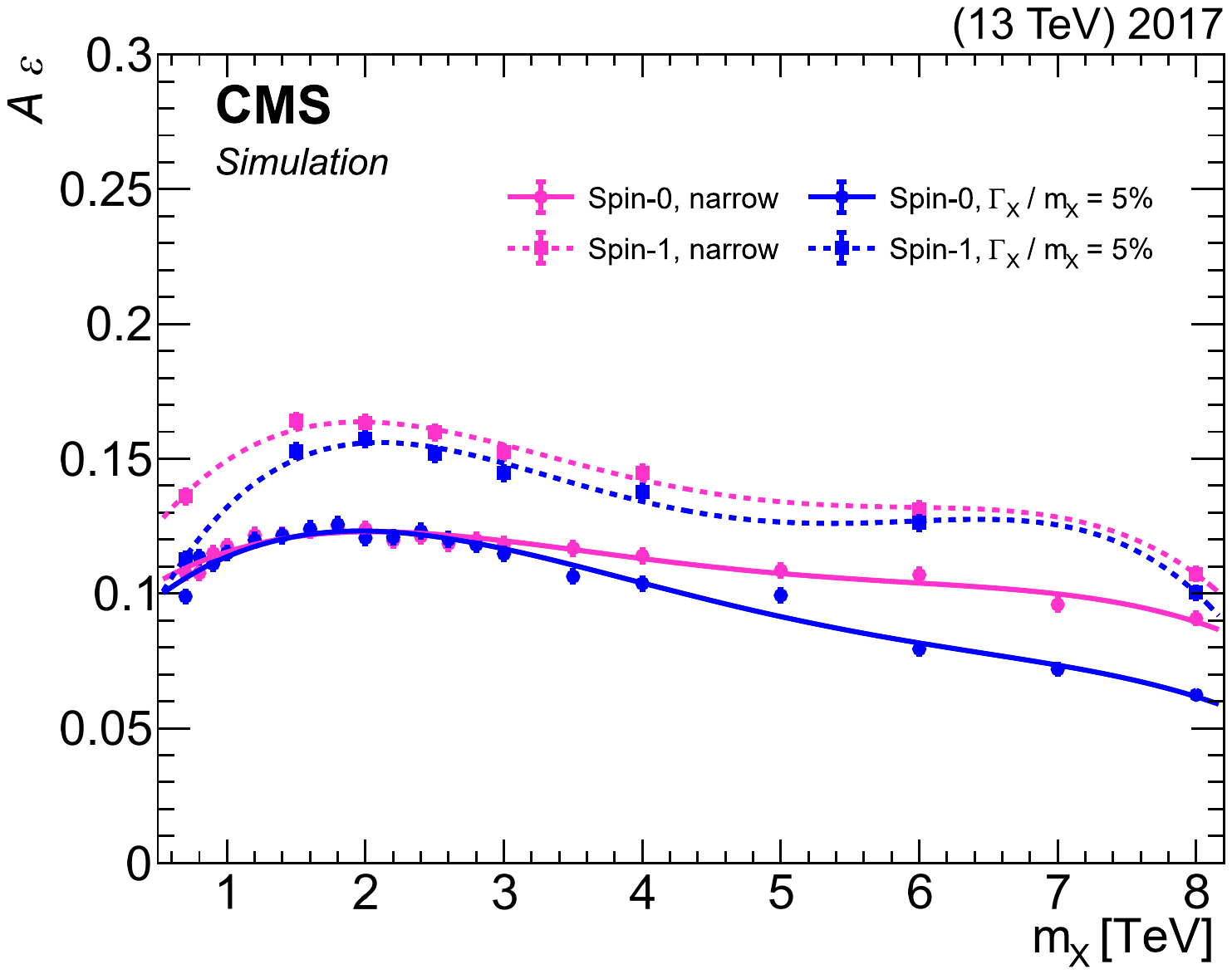}
\includegraphics[width=0.49\textwidth]{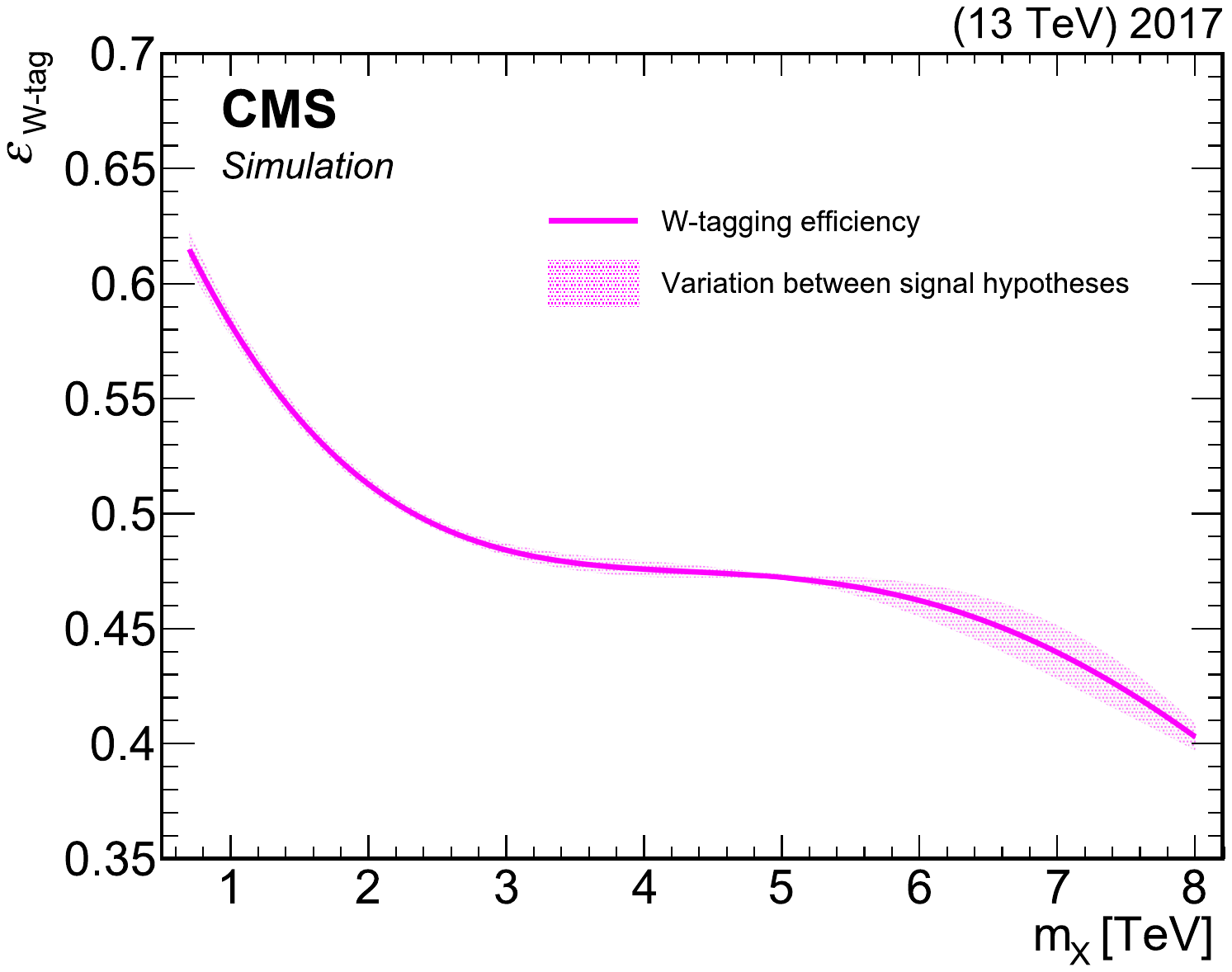}
\caption{Signal acceptance $\mathcal{A}$ (upper left), the product of the signal acceptance and selection efficiency $\mathcal{A}\varepsilon$ (upper right), and the \PW tagging efficiency (lower) for spin-0 (solid lines) and spin-1 (dashed lines) resonances, for the narrow (pink) and broad (blue) hypotheses. The curves are obtained by fitting fourth-order polynomials to the set of discrete mass points, for which simulated signal samples are available. For the \PW tagging efficiency, the average value obtained for the different spin and width hypotheses and the spread of the individual efficiencies about the average are shown with the solid line and the shaded band, respectively.}
\label{fig:eff}
\end{figure*}
\end{linenomath}

After the final selection, the background shape of the \MWg spectrum in the SR is modeled by a background-only fit with a smooth, monotonically falling function. A variety of functional forms are considered for the background fit, and for each function, a goodness-of-fit (GOF) test based on the Kolmogorov--Smirnov statistic is performed in the SR. The nominal background fit function is chosen as the one with the best GOF achieved with the minimal number of parameters:
\begin{linenomath}
\begin{equation}
\frac{\rd N}{\rd m} = p_{0}(m/\sqrt{s})^{p_{1}+p_{2}\log(m/\sqrt{s})+p_{3}\log^{2}(m/\sqrt{s})}
\end{equation}
\end{linenomath}
where $p_{i}$ (with $i =$ 0--3) are the free parameters of the fit. The best fit to data with the background-only hypothesis is shown in Fig.~\ref{fig:data}. This smooth background function provides an adequate description of the data in the entire mass range probed.
\begin{linenomath}
\begin{figure}[htb!]
\centering
\includegraphics[width=0.49\textwidth]{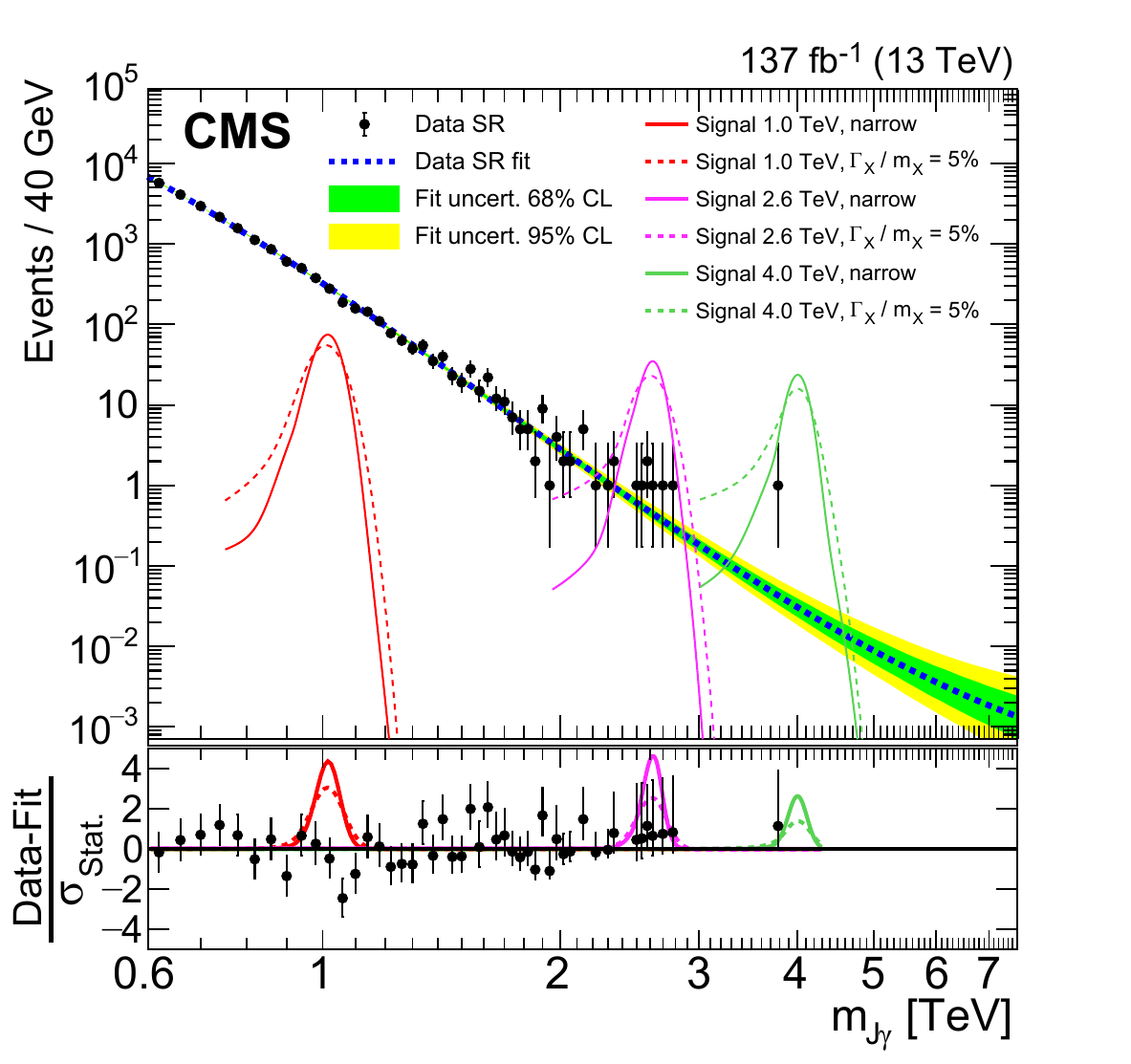}
\caption{Background-only fit to data (black points) with the chosen background function. The green (inner) and yellow (outer) bands show, respectively, the 68 and 95\% confidence level statistical uncertainties in the fit. The lower panel contains the pull distribution, defined as the difference between the data yield and the background prediction, divided by their combined uncertainty. Expected signal shapes are also shown in the lower panel for three different resonance mass hypotheses, 1.0\TeV (red), 2.6\TeV (magenta), and 4.0\TeV (green), and for both the narrow (solid) and broad (dashed) cases. Signal normalizations are set to 15, 1.0, and 0.30\unit{fb}, respectively, for illustrative purposes.}
\label{fig:data}
\end{figure}
\end{linenomath}

\section{Systematic uncertainties}
\label{sec:systematics}

In order to prove that no systematic bias arises from the choice of the background fit function, an alternative fit function that performed relatively well in the GOF test is used to generate a large number of \MWg spectra, with or without signal injection. The spectra are then fit to the sum of the chosen background function and a signal template with the mass and normalization allowed to float. The signal significance is extracted from each fit and the distributions of the pull of the signal yield are constructed, where the pull is defined as the difference between the injected and extracted signal normalizations, divided by the statistical uncertainty in the extracted signal normalization from the fit. We observe that the distributions of the pulls are consistent with a Gaussian function with a zero mean and a standard deviation of unity, and thus conclude that any systematic bias from the background fitting procedure is negligible compared to the statistical uncertainties in the fit. We therefore use the latter as the only uncertainties associated with the background estimate.

The systematic uncertainty associated with the background shape is evaluated via likelihood profiling. This procedure refits for the optimal values of the background parameters for each signal mass hypothesis with the parameters of the background function allowed to vary freely in the fit, thus accounting for the uncertainty in the background prediction.

There are several sources of uncertainties in the description of the signal. Most of the uncertainties mainly affect the signal yield and have only a small impact on the shape of the signal mass distribution. These include the integrated luminosity, trigger plateau efficiency, photon identification efficiency, pileup description, choice of PDFs, and the efficiency of tagging the \PW\ boson jet, including the efficiencies of the SR selection on the \MJ and of the $\tauto$ requirement. Uncertainties affecting both the yield and the shape of the signal distributions include the jet and photon energy scales and resolutions.

The uncertainties in the integrated luminosity are 2.5\%~\cite{LUM-17-001}, 2.3\%~\cite{LUM-17-004}, and 2.5\%~\cite{LUM-18-001} in 2016, 2017, and 2018, respectively. The efficiency of the trigger is obtained using an independent suite of triggers. A systematic uncertainty of 1.0 (2.3)\% is assigned to the trigger efficiency in 2016 and 2018 (2017) data taking, based on the difference between the observed plateau value and unity. The uncertainty due to the photon identification efficiency is obtained by comparing the efficiency in simulation with that in $\PZ \to \Pe\Pe$ events, where electrons are reconstructed as photons. It amounts to a 3--6\% uncertainty in the signal yield.

The uncertainty due to the description of pileup in the simulated samples is estimated by changing the value of the total inelastic cross section by $\pm 4.6\%$~\cite{Sirunyan:2018nqx} and recalculating the signal efficiency after the corresponding change in the pileup distribution. The resulting uncertainty is 1.0--1.5 (1.0--2.0)\% for narrow (broad) resonances. The PDF uncertainty is determined using the PDF4LHC prescription~\cite{Butterworth:2015oua}. Only the effect on the signal acceptance is included as a source of the experimental uncertainty, and amounts to 2\%. The uncertainty in the \PW jet tagging efficiency mainly originates from the $\tauto$ selection efficiency. It amounts to an uncertainty of 3.2--11\% in the signal yield.

The uncertainty in the jet energy scale is obtained by varying the energy scale of the jet corresponding to the \PW candidate by a $\pt$- and $\eta$-dependent correction~\cite{Khachatryan:2016kdb}. The uncertainty due to the jet energy resolution is obtained by smearing the momentum of the \PW jet in simulation to match that in data. The combined effect on the signal yield is 1.3\%, which is the major source of the uncertainty in the signal shapes. The effects of the photon energy scale and resolution are accounted for in a similar fashion; they are significantly less important than those stemming from the jets, because of a much higher precision of the photon energy reconstruction~\cite{CMS-EGM-14-001}.

Given that the signal is extracted from the \MWg spectrum combined over the three data-taking periods, we take into account correlations between the uncertainties across the three years, and use the luminosity-weighted linear (quadratic) average for correlated (uncorrelated) uncertainties. The integrated luminosity uncertainty has both correlated and uncorrelated components, resulting in the overall uncertainty of 1.8\% when applied to the full data set. The trigger and \PW tagging efficiencies, as well as the jet energy scale and resolution uncertainties, are treated as uncorrelated across the three years, while the rest are treated as fully correlated. A summary of the systematic uncertainties, as well as the effect of the year-to-year correlations, is given in Table~\ref{table:syst}.

\begin{linenomath}
\begin{table*}[htb!]
\centering
\topcaption{Systematic uncertainties affecting the signal description. Uncertainties marked with ``$\dagger$" affect both the yield and the shape of the signal distribution, while the rest only affect the signal yield. In cases where the uncertainty is different for various data-taking periods, the three numbers given in the second column correspond to the 2016/2017/2018 data taking, while the third column shows the combined uncertainties across the three years, taking into account the year-to-year correlations. The effect on the signal yield is the same for all the signal hypotheses studied.}
\label{table:syst}
\begin{tabular}{lcc}
\hline
Source & Effect on the signal yield (\%) & Combined (\%)\\[\cmsTabSkip]
\hline
Integrated luminosity & 2.5/2.3/2.5 & 1.8\\
Trigger efficiency & 1.0/2.3/1.0 & 0.9\\
Photon ident. efficiency & 4.7/6.0/3.0 & 4.4\\
Pileup & 1.0/2.0/1.0 & 1.3\\
PDF & 2.0 & 2.0\\
\PW tagging efficiency & 11/7.4/3.2 & 3.9\\
Jet energy scale and resolution$^\dagger$ & 1.3 & 0.8\\
Photon energy scale and resolution$^\dagger$ & 0.5/1.0/1.0 & 0.9\\[\cmsTabSkip]

Total & 12.6/10.6/5.8 & 6.7\\
\hline
\end{tabular}
\end{table*}
\end{linenomath}

\section{Results}
\label{sec:results}

We set model-specific upper limits on the product of the cross section and branching fraction for both narrow and broad, spin-0 and spin-1 resonances using the modified frequentist \CLs criterion~\cite{Junk:1999kv,Read:2002hq,ATL-PHYS-PUB-2011-011}, with a likelihood ratio in the asymptotic approximation~\cite{Cowan} used as a test statistic. The yield (shape) uncertainties are incorporated as nuisance parameters with log-normal (Gaussian) priors. These limits, at 95\% \CL, are shown in Fig.~\ref{fig:limits}, separately for the spin-0 and spin-1 hypotheses, as well as for narrow and broad resonances, and are the most restrictive limits to date on the existence of \Wg resonances over the majority of the masses probed. The $p$-values for the background-only fit are shown in Fig.~\ref{fig:p-values} for the narrow (left) and broad (right) resonances for both spin hypotheses. The largest excess seen in the limit plots has a mass around 1.58\TeV, with a local significance of 2.8 (3.1) standard deviations for narrow (broad) signals for both spin hypotheses. After taking into account the look-elsewhere effect~\cite{LEE}, the global significance of the excess is estimated to be 1.1 (1.7) standard deviations, favoring its interpretation as a statistical fluctuation in data. 

In the case of narrow resonances, we compare these limits with the predictions of two of the models described in Ref.~\cite{Capdevilla:2019zbx}, which we use as benchmarks. In the spin-0 case, a scalar or pseudoscalar $SU(2)_\mathrm{L}$ triplet $\phi^\alpha$ that couples to the SM vector boson fields via anomaly-induced interactions $\phi^aW_{\mu\nu}^a \widetilde B^{\mu\nu}/\Lambda$ and to the SM fermionic fields with an effective coupling $y_m/\Lambda$, is considered. Here, $\Lambda$ is the ultraviolet cutoff of the model, chosen to be 2, 4, or 5 times the resonance mass, while $y_m$ is set to 0.10 or 0.15 to suppress fermionic decays. In the spin-1 case, a vector $SU(2)_\mathrm{L}$ triplet $V^a_\mu$ that couples to the SM vector boson fields via a higher-dimensional operator $c_WV^{a\nu}_\mu W^{a\alpha}_\nu B^\mu_\alpha/\Lambda^2$ and to the SM fermionic fields directly with the coupling $g_m$ is considered. Similar to the scalar triplet case, we set $\Lambda$ to 4 or 5 times the resonance mass, $c_W$ to 1, and $g_m$ to 0.10 or 0.15. Additionally, we set the coupling of the vector triplet to the Higgs boson $c_h$ to zero. These choices of parameters assure narrow resonances in the mass range of interest, with sizable branching fractions to \Wg. As a result of this search, benchmark heavy scalar (vector) triplet bosons with masses between 0.75 (1.15) and 1.40 (1.36)\TeV are excluded at 95\% \CL, as shown in Fig.~\ref{fig:limits}. Spin-0 $\pi_3$ states of Ref.~\cite{Arkani-Hamed:2016kpz} are beyond the sensitivity of this search, as they decay predominantly into hadronic final states and the branching fraction into the \Wg channel is suppressed.

The asymptotic \CLs approach tends to overestimate the limits in the bins with low event count. For the narrow (broad) spin-0 signal hypothesis, it is found that the asymptotic approximation yields more stringent cross section limits compared to a full \CLs approach above 2.6 (3.2)\TeV. The limits are 10 (16)\% more stringent at 3.5\TeV and the difference increases to 31\% at 6.0\TeV in both cases. Similar behavior is expected for the spin-1 signal hypothesis. The asymptotic approximation has negligible impact on the mass exclusions in the benchmark models, as these limits are well below 2.6\TeV.

\begin{linenomath}
\begin{figure*}[htb!]
\centering
\includegraphics[width=0.48\textwidth]{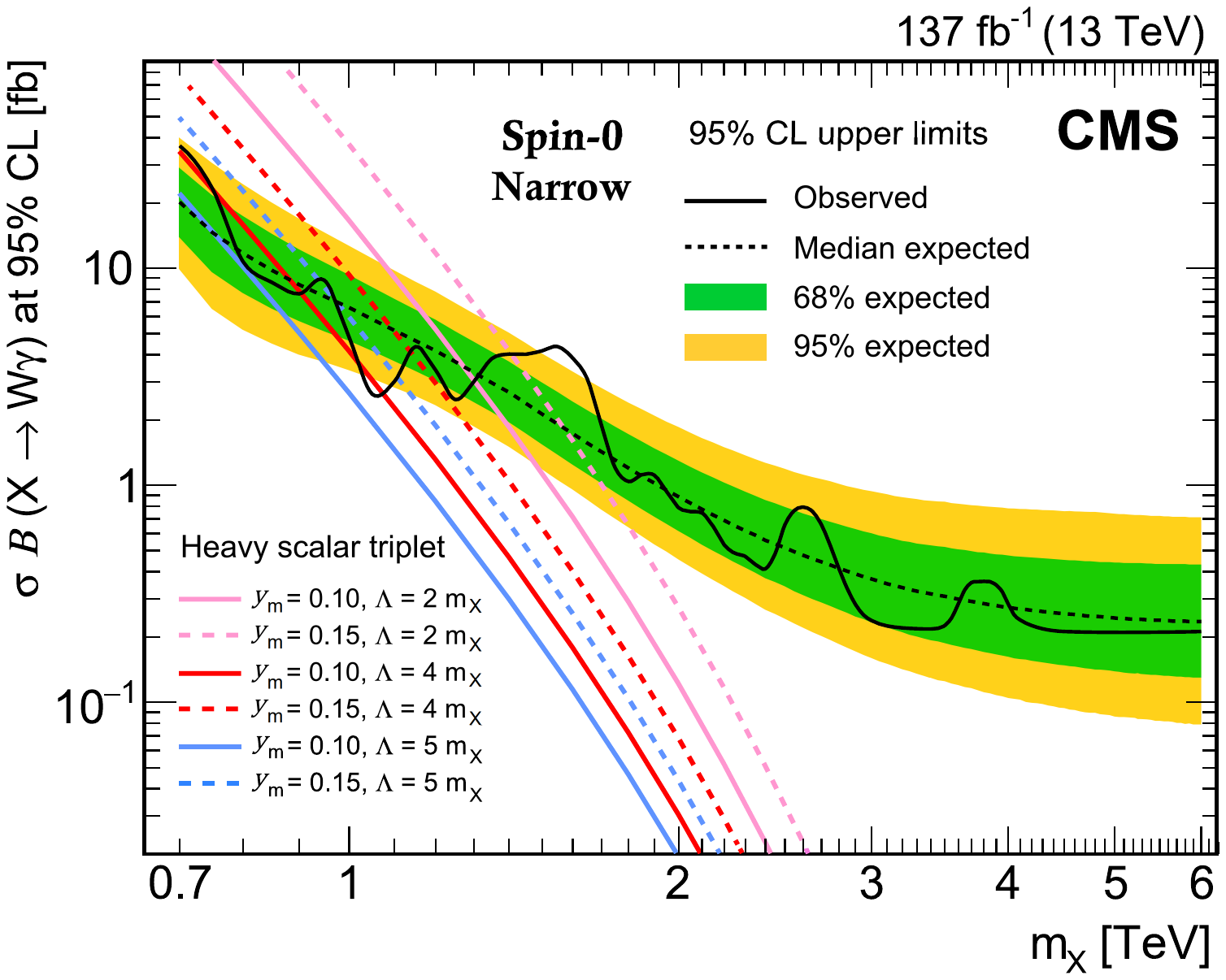}
\includegraphics[width=0.48\textwidth]{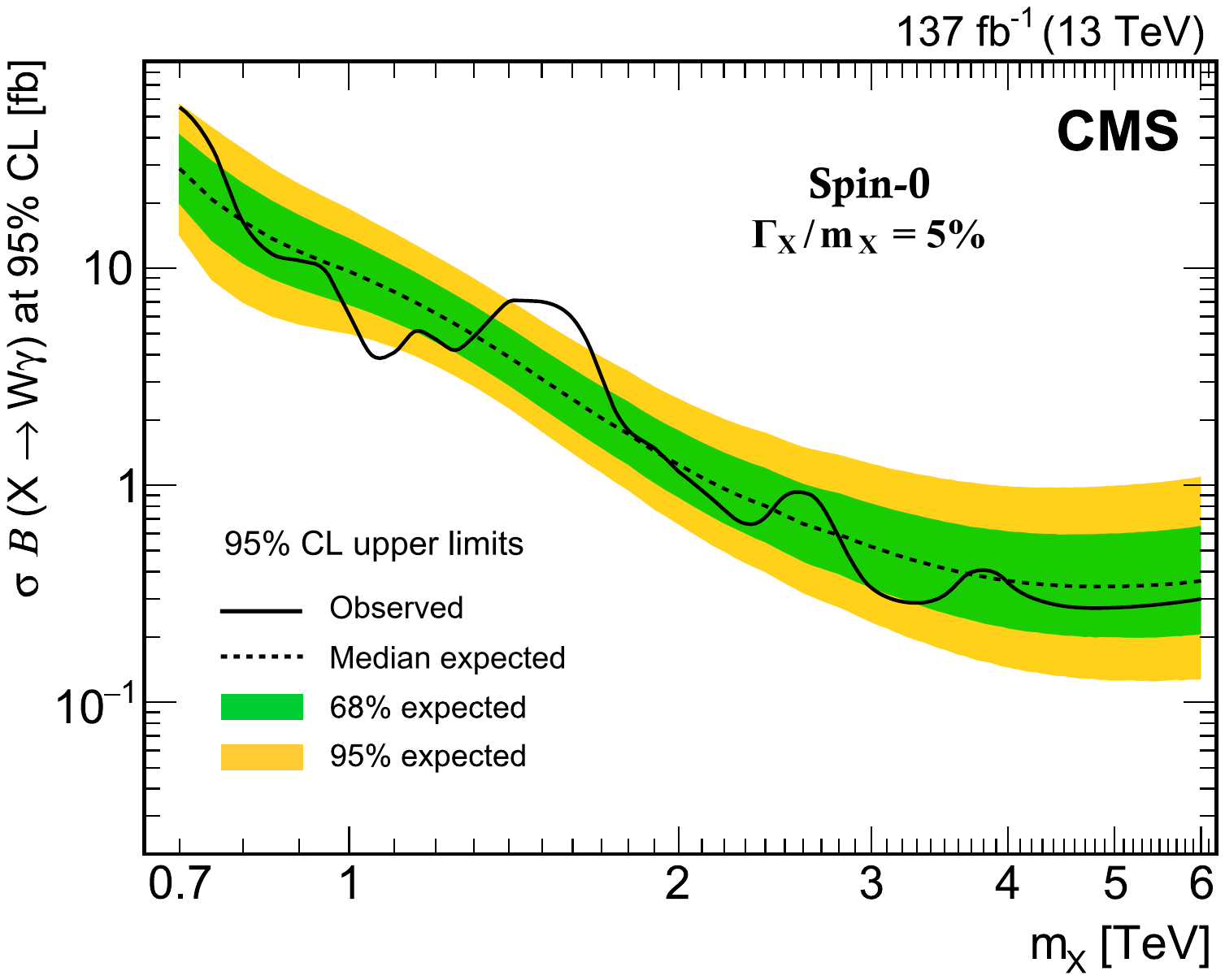}
\includegraphics[width=0.48\textwidth]{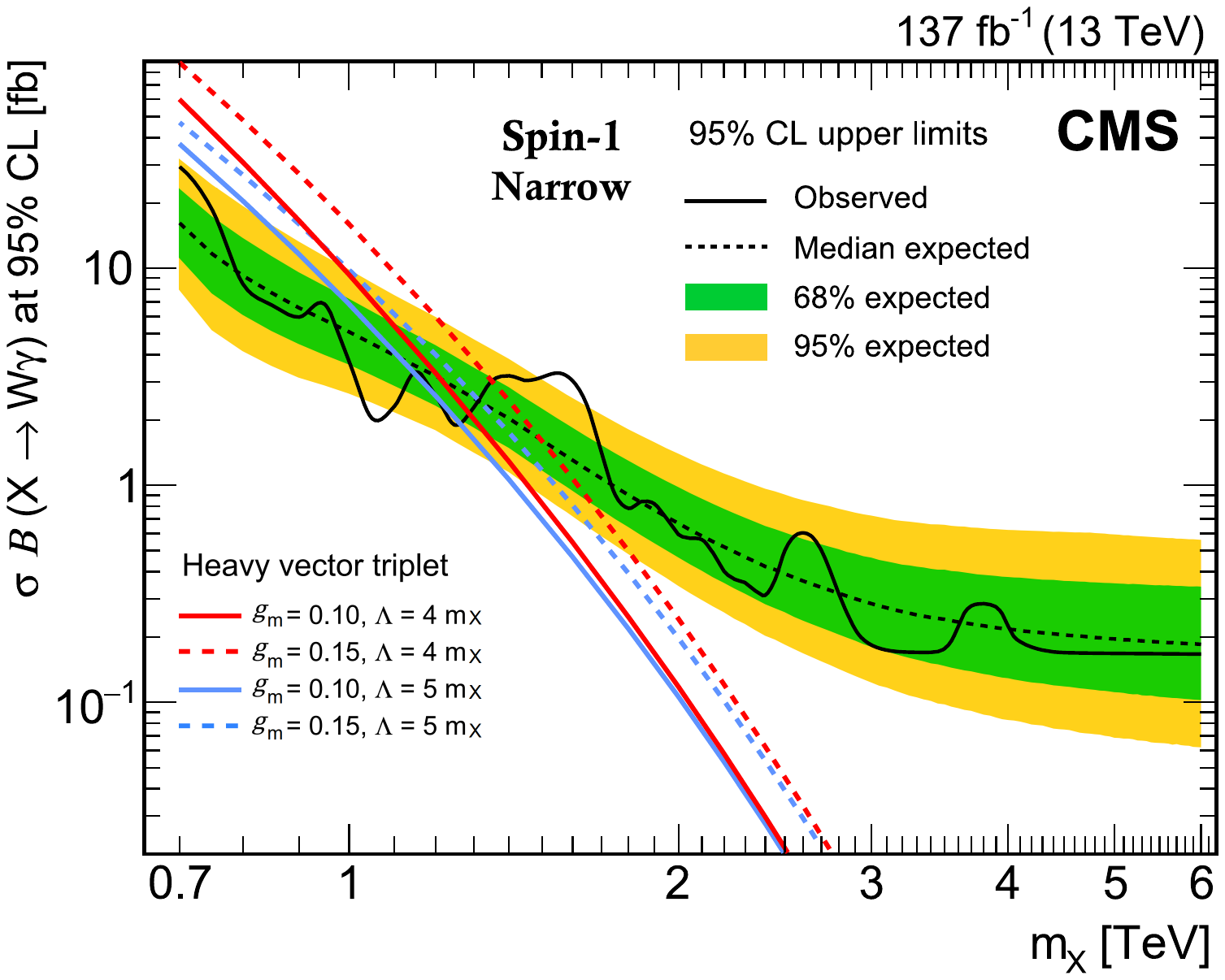}
\includegraphics[width=0.48\textwidth]{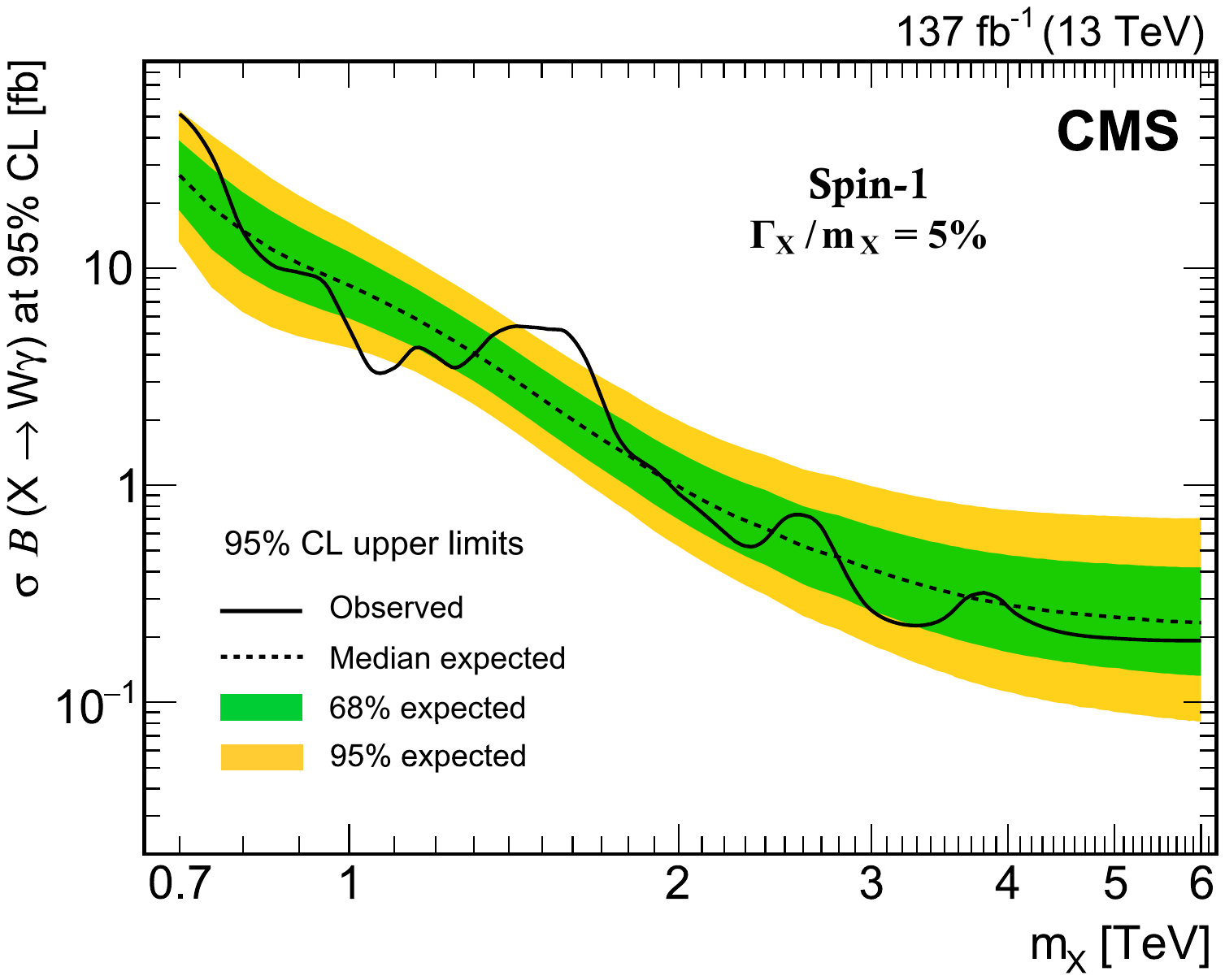}
\caption{Expected and observed 95\% confidence level limits on $\sigma \mathcal{B}(\PX \to \Wg)$ for the spin-0 (upper row) and spin-1 (lower row) resonances for the narrow (left column) and broad (right column) resonance cases. Also shown, for spin-0 (spin-1) narrow-resonance case, theoretical cross sections for heavy scalar (vector) triplet resonance production in the benchmark model of Ref.~\protect\cite{Capdevilla:2019zbx}, which can be probed by this search.}
\label{fig:limits}
\end{figure*}
\end{linenomath}

\begin{linenomath}
\begin{figure*}[htb!]
\centering
\includegraphics[width=0.48\textwidth]{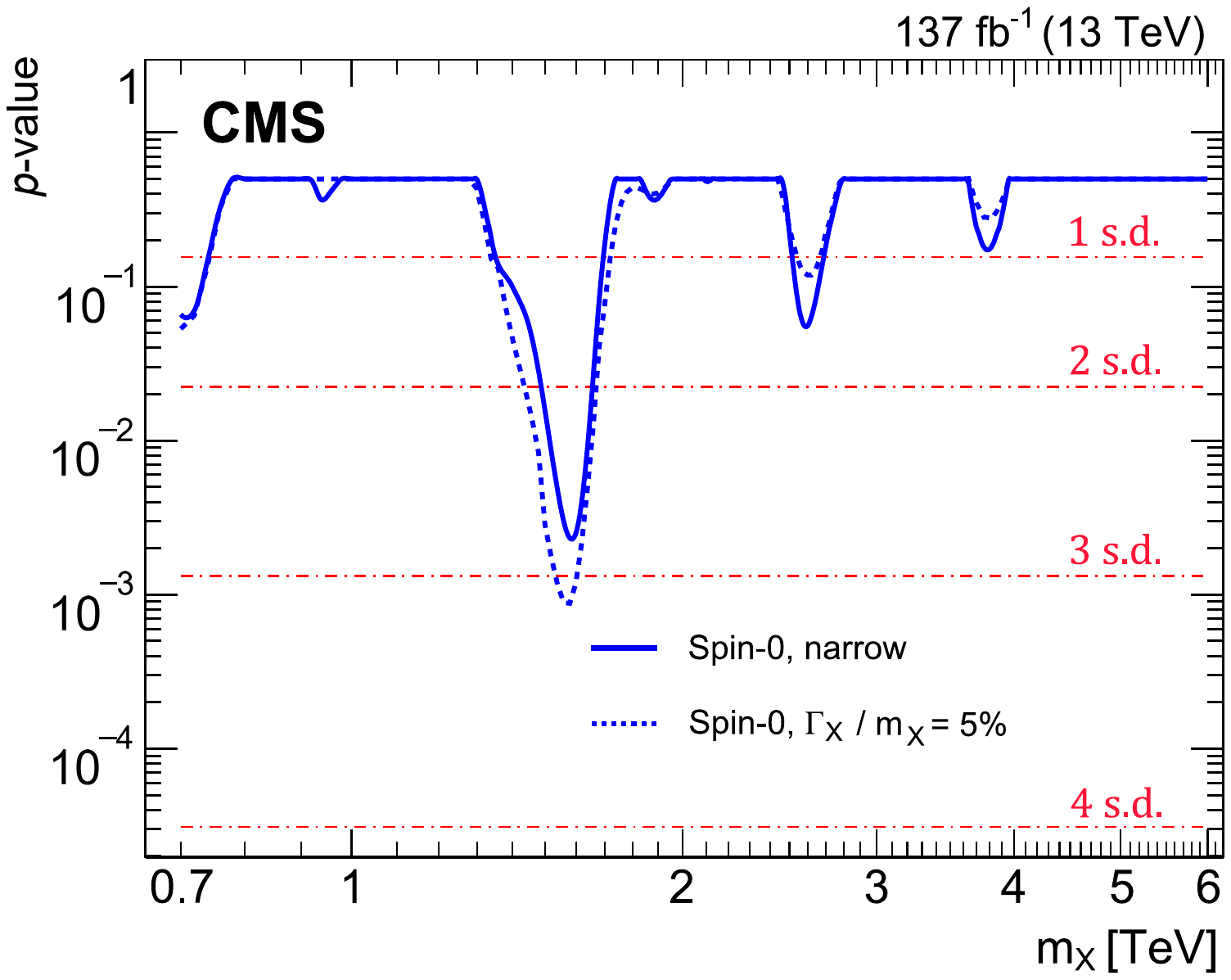}
\includegraphics[width=0.48\textwidth]{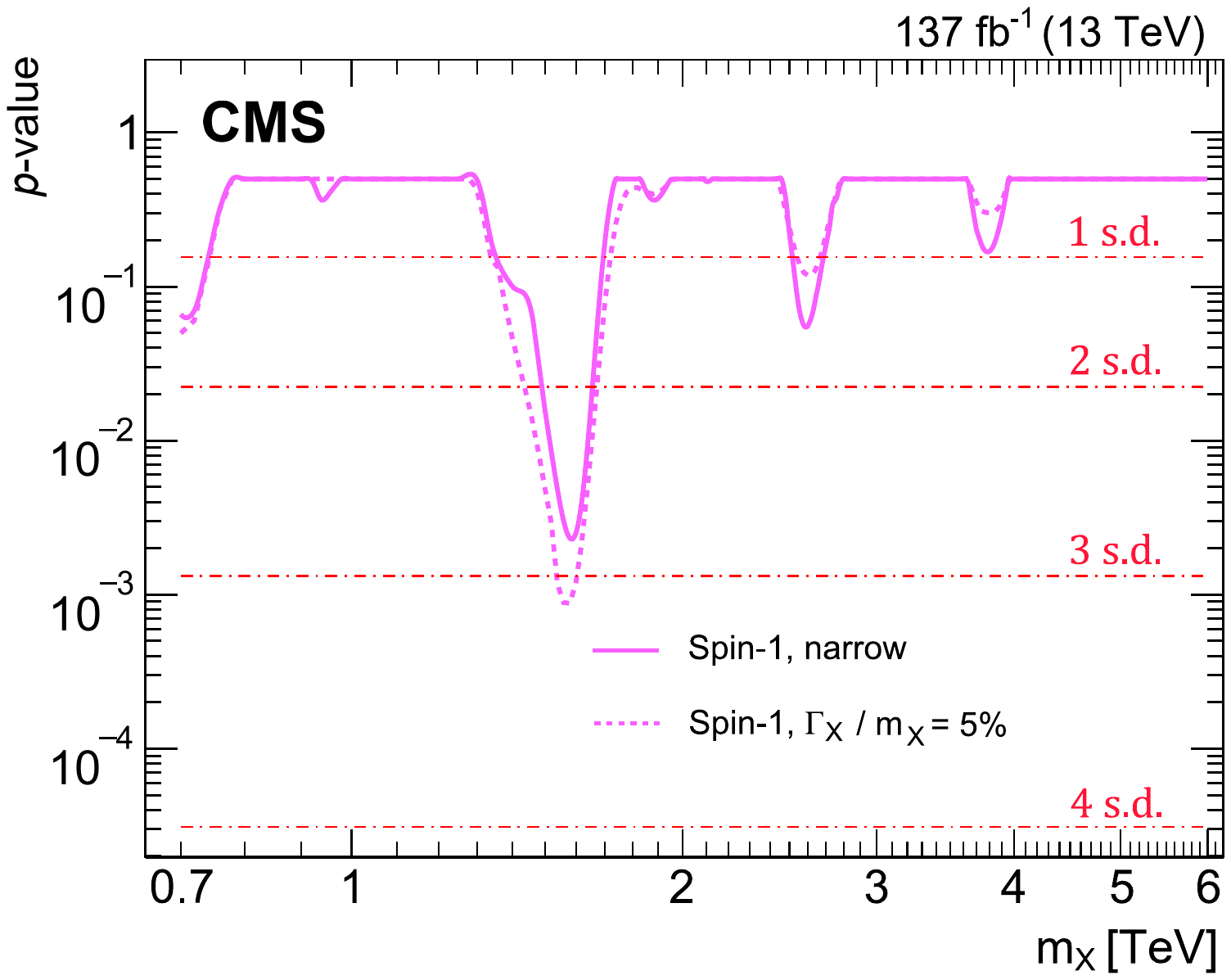}
\caption{Observed local $p$-values for spin-0 (left) and spin-1 (right) resonance hypotheses. The largest excess observed at 1.58\TeV corresponds to a local significance of 2.8 (3.1) standard deviations (s.d.) for narrow (broad) signals, for both spin hypotheses.}
\label{fig:p-values}
\end{figure*}
\end{linenomath}

\begin{linenomath}
\begin{figure*}[htb!]
\centering
\includegraphics[width=0.48\textwidth]{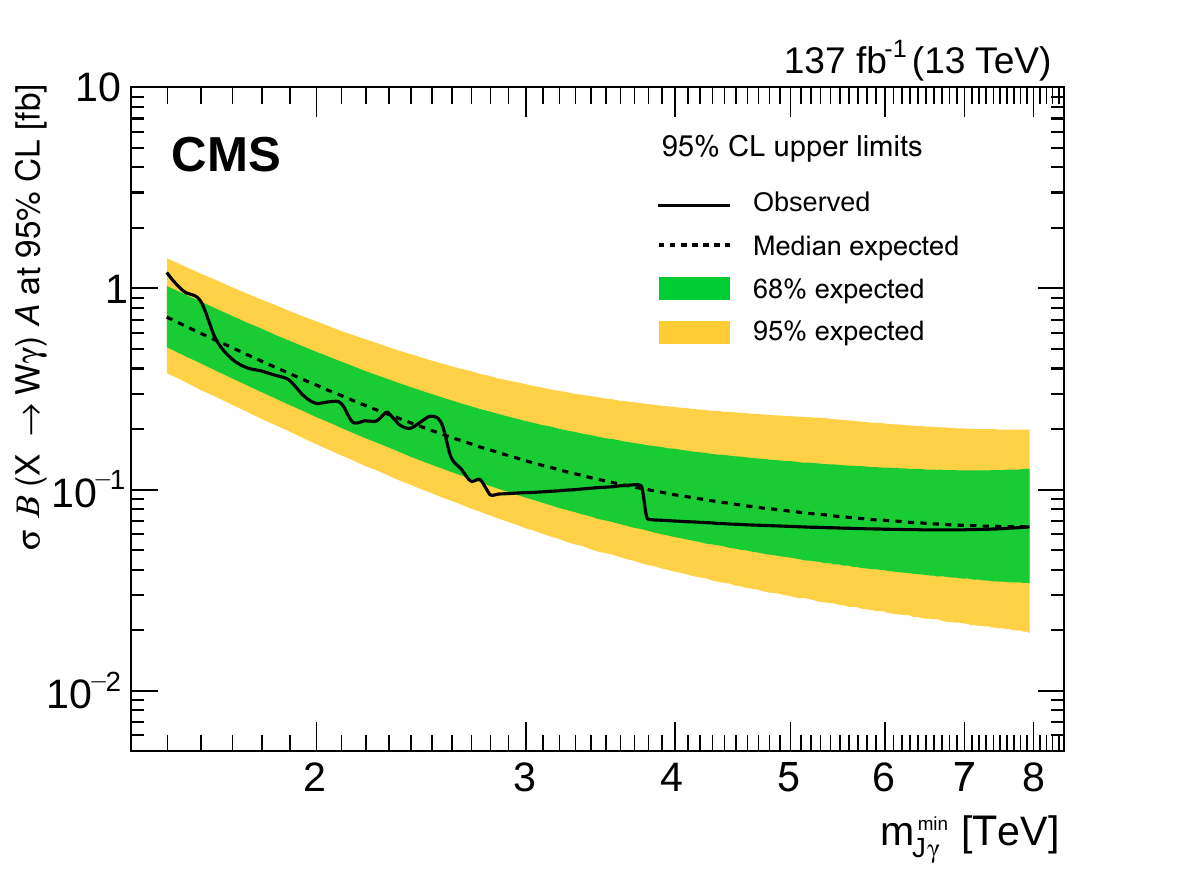}
\includegraphics[width=0.48\textwidth]{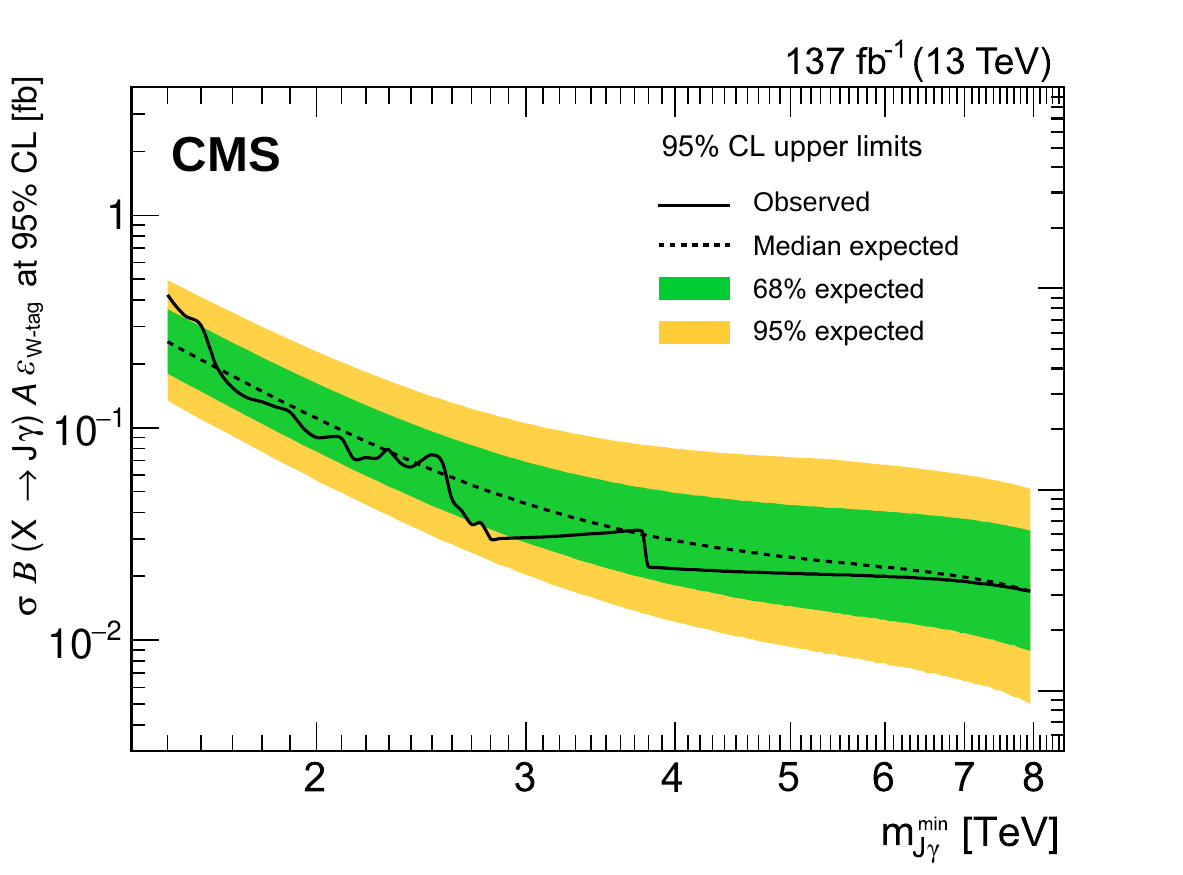}
\caption{Expected and observed 95\% confidence level model-independent limits on $\sigma \mathcal{B}(\PX \to \Wg) \mathcal{A}$ (left) and $\sigma \mathcal{B}(\PX \to \Jg) \mathcal{A} \epsilon_{\PW\text{-tag}}$ (right), as a function of the minimum invariant mass requirement on the \Jg system.}
\label{fig:limits_counting}
\end{figure*}
\end{linenomath}

Model-independent upper limits on the product of the cross section, branching fraction, and signal acceptance are set in the context of a simple counting experiment that considers the number of events observed and expected above a variable \Jg invariant mass threshold. These limits allow the interpretation of our results in nonresonant and other resonant models by estimating their acceptance for the selections used in this analysis. Since the exact description of the signal shape is no longer important in the region where little background is expected, these limits are continued up to the mass threshold of 8.0 TeV, thus extending the mass range of phenomena that can be investigated. To make this interpretation possible for a broad set of models, we treat the \PW tagging efficiency as a part of either the experimental efficiency or the signal acceptance. The former case, shown in Fig.~\ref{fig:limits_counting} (left), would apply to models yielding signatures with a hadronically decaying Lorentz-boosted \PW boson, as the \PW tagging efficiency in this case should be very similar to that in the model-specific analysis. The latter case, shown in Fig.~\ref{fig:limits_counting} (right), would apply to an even broader set of models, \eg, the ones predicting ${\PZ}{\PGg}$ signatures, by accounting for the efficiency of tagging a \PZ boson using our \PW tagging requirements as a part of the signal acceptance.

\section{Summary}
\label{sec:summary}

A search for \Wg resonances in the mass range between 0.7 and 6.0\TeV has been presented. The \PW\ boson is reconstructed from its hadronic decay, in which the final-state products form a single large-radius jet owing to the large Lorentz boost of the \PW\ boson. The search is based on proton-proton collision data collected at $\sqrt{s} = 13\TeV$ with the CMS detector at the LHC in 2016--2018, corresponding to an integrated luminosity of 137\fbinv. No significant excess above the smoothly falling background is observed. Limits at 95\% confidence level on the product of the cross section and branching fraction for \Wg resonances are set, ranging from 37 (55) to 0.21 (0.30)\unit{fb} for the narrow (broad) spin-0 hypothesis, and from 29 (51) to 0.17 (0.19)\unit{fb} for the narrow (broad) spin-1 hypothesis. The results reported are the most restrictive limits to date on the existence of such resonances over a large range of probed masses. In specific narrow-resonance benchmark models, heavy scalar (vector) triplet resonances with masses between 0.75 (1.15) and 1.40 (1.35)\TeV are excluded for a range of model parameters probed. In addition, model-independent limits are set on the product of the cross section, branching fraction, and signal acceptance, as functions of the minimum invariant mass of the jet-photon system, making possible the interpretation of these results in the context of a broader class of models predicting similar signatures.

\begin{acknowledgments}
    We congratulate our colleagues in the CERN accelerator departments for the excellent performance of the LHC and thank the technical and administrative staffs at CERN and at other CMS institutes for their contributions to the success of the CMS effort. In addition, we gratefully acknowledge the computing centers and personnel of the Worldwide LHC Computing Grid and other centers for delivering so effectively the computing infrastructure essential to our analyses. Finally, we acknowledge the enduring support for the construction and operation of the LHC, the CMS detector, and the supporting computing infrastructure provided by the following funding agencies: BMBWF and FWF (Austria); FNRS and FWO (Belgium); CNPq, CAPES, FAPERJ, FAPERGS, and FAPESP (Brazil); MES (Bulgaria); CERN; CAS, MoST, and NSFC (China); MINCIENCIAS (Colombia); MSES and CSF (Croatia); RIF (Cyprus); SENESCYT (Ecuador); MoER, ERC PUT and ERDF (Estonia); Academy of Finland, MEC, and HIP (Finland); CEA and CNRS/IN2P3 (France); BMBF, DFG, and HGF (Germany); GSRT (Greece); NKFIA (Hungary); DAE and DST (India); IPM (Iran); SFI (Ireland); INFN (Italy); MSIP and NRF (Republic of Korea); MES (Latvia); LAS (Lithuania); MOE and UM (Malaysia); BUAP, CINVESTAV, CONACYT, LNS, SEP, and UASLP-FAI (Mexico); MOS (Montenegro); MBIE (New Zealand); PAEC (Pakistan); MSHE and NSC (Poland); FCT (Portugal); JINR (Dubna); MON, RosAtom, RAS, RFBR, and NRC KI (Russia); MESTD (Serbia); SEIDI, CPAN, PCTI, and FEDER (Spain); MOSTR (Sri Lanka); Swiss Funding Agencies (Switzerland); MST (Taipei); ThEPCenter, IPST, STAR, and NSTDA (Thailand); TUBITAK and TAEK (Turkey); NASU (Ukraine); STFC (United Kingdom); DOE and NSF (USA).
     
    \hyphenation{Rachada-pisek} Individuals have received support from the Marie-Curie program and the European Research Council and Horizon 2020 Grant, contract Nos.\ 675440, 724704, 752730, 765710 and 824093 (European Union); the Leventis Foundation; the Alfred P.\ Sloan Foundation; the Alexander von Humboldt Foundation; the Belgian Federal Science Policy Office; the Fonds pour la Formation \`a la Recherche dans l'Industrie et dans l'Agriculture (FRIA-Belgium); the Agentschap voor Innovatie door Wetenschap en Technologie (IWT-Belgium); the F.R.S.-FNRS and FWO (Belgium) under the ``Excellence of Science -- EOS" -- be.h project n.\ 30820817; the Beijing Municipal Science \& Technology Commission, No. Z191100007219010; the Ministry of Education, Youth and Sports (MEYS) of the Czech Republic; the Deutsche Forschungsgemeinschaft (DFG), under Germany's Excellence Strategy -- EXC 2121 ``Quantum Universe" -- 390833306, and under project number 400140256 - GRK2497; the Lend\"ulet (``Momentum") Program and the J\'anos Bolyai Research Scholarship of the Hungarian Academy of Sciences, the New National Excellence Program \'UNKP, the NKFIA research grants 123842, 123959, 124845, 124850, 125105, 128713, 128786, and 129058 (Hungary); the Council of Science and Industrial Research, India; the Latvian Council of Science; the Ministry of Science and Higher Education and the National Science Center, contracts Opus 2014/15/B/ST2/03998 and 2015/19/B/ST2/02861 (Poland); the National Priorities Research Program by Qatar National Research Fund; the Ministry of Science and Higher Education, project no. 0723-2020-0041 (Russia); the Programa Estatal de Fomento de la Investigaci{\'o}n Cient{\'i}fica y T{\'e}cnica de Excelencia Mar\'{\i}a de Maeztu, grant MDM-2015-0509 and the Programa Severo Ochoa del Principado de Asturias; the Thalis and Aristeia programs cofinanced by EU-ESF and the Greek NSRF; the Rachadapisek Sompot Fund for Postdoctoral Fellowship, Chulalongkorn University and the Chulalongkorn Academic into Its 2nd Century Project Advancement Project (Thailand); the Kavli Foundation; the Nvidia Corporation; the SuperMicro Corporation; the Welch Foundation, contract C-1845; and the Weston Havens Foundation (USA).
\end{acknowledgments}

\bibliography{auto_generated} 
\cleardoublepage \appendix\section{The CMS Collaboration \label{app:collab}}\begin{sloppypar}\hyphenpenalty=5000\widowpenalty=500\clubpenalty=5000\vskip\cmsinstskip
\textbf{Yerevan Physics Institute, Yerevan, Armenia}\\*[0pt]
A.~Tumasyan
\vskip\cmsinstskip
\textbf{Institut f\"{u}r Hochenergiephysik, Wien, Austria}\\*[0pt]
W.~Adam, J.W.~Andrejkovic, T.~Bergauer, S.~Chatterjee, M.~Dragicevic, A.~Escalante~Del~Valle, R.~Fr\"{u}hwirth\cmsAuthorMark{1}, M.~Jeitler\cmsAuthorMark{1}, N.~Krammer, L.~Lechner, D.~Liko, I.~Mikulec, F.M.~Pitters, J.~Schieck\cmsAuthorMark{1}, R.~Sch\"{o}fbeck, M.~Spanring, S.~Templ, W.~Waltenberger, C.-E.~Wulz\cmsAuthorMark{1}
% ...
\vskip\cmsinstskip
% ...
1:  Also at TU Wien, Wien, Austria\\
% ...
\end{sloppypar}
\end{document}